\documentclass[12pt]{article}

\usepackage{a41}

\usepackage{dsfont}

\usepackage{palatino}
\usepackage{color}
\usepackage{amsmath}                
\usepackage{booktabs}
\usepackage{microtype}
\usepackage{cite}
\usepackage{amsmath}
\usepackage{amssymb}
\usepackage{color}
\usepackage[rflt]{floatflt}
\usepackage{float}

\usepackage[rflt]{floatflt}
\usepackage{float}
\usepackage{slashed}
\newmuskip\pFqmuskip

\newcommand*\pFq[6][8]{%
  \begingroup % only local assignments
  \pFqmuskip=#1mu\relax
  \mathchardef\normalcomma=\mathcode`,
  \mathcode`\,=\string"8000
  \begingroup\lccode`\~=`\,
  \lowercase{\endgroup\let~}\pFqcomma
  {}_{#2}F_{#3}{\left[\genfrac..{0pt}{}{#4}{#5};#6\right]}%
  \endgroup
}
\newcommand{\pFqcomma}{{\normalcomma}\mskip\pFqmuskip}

\def\b0{\beta_0}

\usepackage{array}

\usepackage[english]{babel}
\usepackage[latin1]{inputenc}
\usepackage[T1]{fontenc}
\usepackage{ae}

\usepackage{url}

\usepackage{amsmath, amsthm, amssymb}
\newtheorem{thm}{Theorem}[section]

\newtheorem{definition}[thm]{Definition}

\newcommand{\ep}{\varepsilon}

\usepackage{rotating}

\usepackage{graphicx}
\newcounter{mmacnt}
\def\restartmma{\setcounter{mmacnt}{0}}
\restartmma \catcode`|=\active
\def|#1|{\mathrm{#1}}
\catcode`|=12
\newenvironment{mma}{
 \par\smallskip
 \catcode`|=\active
 \parskip=0pt\parindent=0pt % locally
 \small
 \def\In##1\\{%
\def\linebreak{\hfill\break\null\qquad}%
\refstepcounter{mmacnt}
\hangindent=2.5em\hangafter=0
\leavevmode
\llap{\tiny\sffamily n[\arabic{mmacnt}]:=\kern.5em}%
\mathversion{bold}\footnotesize$\displaystyle##1$\normalsize
\mathversion{normal}\par
 }%
 \def\Print##1\\{%
\def\linebreak{\hfill\break}%
\hangindent=2.5em\hangafter=0
\leavevmode ##1\par}%
 \def\Out##1\\{%
\def\linebreak{$\hfill\break\null\hfill$}%
\kern\abovedisplayskip\par
\hangindent=2.5em\hangafter=0
\leavevmode
\llap{\tiny\sffamily Out[\arabic{mmacnt}]=\kern.5em}
\footnotesize$\displaystyle##1$\normalsize\hfill\null\par
\kern\belowdisplayskip
 }%
 \def\Warning##1##2\\{%
\def\linebreak{\hfill\break}%
\hangindent=2.5em\hangafter=0
\leavevmode
{\scriptsize##1 : ##2}\par}%
}{%
 \par\smallskip
}

\usepackage{color}

\newenvironment{fshaded}{%
\MakeFramed {\FrameRestore}
}%
{\endMakeFramed}

\def\b0{\beta_0}

\def\Gp0{{\Gamma^{'}_0}}
\def\Gp1{{\Gamma^{'}_1}}
\def\Gp2{{\Gamma^{'}_2}}

\allowdisplaybreaks[4]

\begin{document}
\setlength{\baselineskip}{0.515cm}

\sloppy
\thispagestyle{empty}
\begin{flushleft}
DESY 20--062
%%\hfill {\tt arXiv:2003.01745 [gr-qc]}
\\
DO--TH 20/04\\
SAGEX--20--10\\
October 2020\\
\end{flushleft}

\mbox{}
\vspace*{\fill}
\begin{center}

{\Large\bf The fifth-order post-Newtonian Hamiltonian dynamics of} 

\vspace*{3mm}
{\Large\bf  two-body systems from an effective field theory approach:}

\vspace*{3mm}
{\Large\bf  potential contributions}

\vspace{3cm} \large
{\large J.~Bl\"umlein$^a$, A.~Maier$^a$, P.~Marquard$^a$, and G.~Sch\"afer$^b$}

\vspace{1.cm}
\normalsize
{\it $^a$Deutsches Elektronen--Synchrotron, DESY,}\\
{\it   Platanenallee 6, D--15738 Zeuthen, Germany}

\vspace*{2mm}
{\it $^b$Theoretisch-Physikalisches Institut, Friedrich Schiller-Universit\"at, \\
Max Wien-Platz 1, D--07743 Jena, Germany}\\

%%\today

\end{center}
\normalsize
\vspace{\fill}
\begin{abstract}
\noindent
We calculate the potential contributions of the motion of binary mass systems in gravity to the fifth 
post--Newtonian order ab initio using coupling and velocity expansions within an effective field theory 
approach based on Feynman amplitudes starting with harmonic coordinates and using dimensional regularization. 
Furthermore, the singular and logarithmic tail contributions are calculated. We also consider the non--local tail 
contributions. Further steps towards the complete calculation are discussed and first comparisons are 
given to results in the literature.
\end{abstract}

\vspace*{\fill}
\noindent
% \numberwithin{equation}{section}
%%%%%%%%%%%%%%%%%%%%%%%%%%%%%%%%%%%%%%%%%%%%%%%%%%%%%%%%%%%%%%%%%%%%%%%%%%%%%%%%%%%%%%%%%%%%%%%%%%%%%%%%%%%%%%%%%%%%%%%%%%%%%%%%%%%
\newpage
%----------------------------------------------------------------------------------------------------------------
\section{Introduction}
\label{sec:1}
%----------------------------------------------------------------------------------------------------------------

\vspace*{1mm}
\noindent
The measurement of gravitational wave signals from merging black holes and neutron stars \cite{LIGO}
has been a recent milestone in astrophysics. The different gravitational wave detectors are reaching
higher and higher sensitivity \cite{PROJECT}, which requests to provide more detailed predictions at the
theoretical side. Currently in binary Hamiltonian dynamics the level of the 4th post--Newtonian (PN) order 
has been fully understood and  agreeing results have been obtained using a variety of different computation
techniques in quite a series of gauges which lead to identical predictions in all key observables 
\cite{Damour:2014jta,FOURTH,Damour:2016abl,Bernard:2016wrg,Damour:2017ced,Foffa:2016rgu,Blumlein:2020pog}. Moreover, 
it has 
been shown by applying canonical transformations \cite{Blumlein:2020pog}, that all descriptions are dynamically 
equivalent. The different approaches can only be compared either by using canonical transformations, which requires
local representations, or by calculating observables.

At the level of the 5th post--Newtonian order, first two agreeing results on the static 
potential
in the harmonic gauge were calculated \cite{Foffa:2019hrb,Blumlein:2019zku}. Later partial results
were derived using different matching techniques for the Hamiltonian in the effective one body (EOB) 
approach in \cite{Bini:2019nra,Bini:2020wpo}.\footnote{First results at 6PN order have been given in 
\cite{Blumlein:2020znm,Bini:2020nsb,Bini:2020hmy} recently. There is also a lot of activity in calculating 
post--Minkowskian corrections, cf.~\cite{Blumlein:2020znm} Ref.~[12],  and 
\cite{Bern:2019nnu,Bern:2019crd,Blumlein:2019bqq,Damour:2019lcq,AccettulliHuber:2020oou,KALIN,Damour:2020tta}.}
Here two parameters, $\bar{d}_5$ and $a_6$, which are of $O(\nu^2)$, with $\nu = m_1 m_2/(m_1+m_2)^2$,
remained yet undetermined.

The conserved Hamiltonian of the motion of binary mass systems in gravity has the following expansion  
%\begin{linenomath*}
%----------------------------------------------------------------------------------------------------
\begin{equation}
H = \sum_{k = 0}^\infty H_{\rm kPN},
\end{equation}
%----------------------------------------------------------------------------------------------------
%\end{linenomath*}
where $k$ labels the post--Newtonian order,\footnote{Here we do not deal with conserved half PN contributions 
occurring from 5.5 PN onward.} with $H_{\rm 0PN} \equiv H_{\rm N}$. From $k = 4$ onward $H_{\rm kPN}$ consists out of 
the term due to potential interactions, $H^{\rm pot}$, and the tail terms, $H^{\rm tail}$,
%\begin{linenomath*}
%----------------------------------------------------------------------------------------------------
\begin{equation}
H_{\rm kPN} = H_{\rm kPN}^{\rm pot} + H_{\rm kPN}^{\rm tail}. 
\end{equation}
%----------------------------------------------------------------------------------------------------
%\end{linenomath*}
In effective field theory approaches\footnote{For the 4PN calculations see 
\cite{Damour:2014jta,FOURTH,Blumlein:2019zku}.} 
based on Feynman diagrams this is the most natural decomposition. In \cite{Bini:2019nra,Bini:2020wpo} another 
decomposition has been chosen into the so--called non--local terms $H_{\rm kPN}^{\rm nl}$ and the local terms 
$H_{\rm kPN}^{\rm loc}$,
%----------------------------------------------------------------------------------------------------
\begin{equation}
H_{\rm kPN} = H_{\rm kPN}^{\rm loc} + H_{\rm kPN}^{\rm nl}. 
\label{eq:lonl}
\end{equation}
%----------------------------------------------------------------------------------------------------
The non--local terms are fully contained in the tail terms and the local contributions are given by the
local parts of the tail terms and the potential contributions.

In the present paper we calculate the 5PN potential corrections and some first parts of the 5PN tail terms 
using an effective field theory (EFT) approach; for related reviews see 
\cite{Goldberger:2007hy,Blanchet:2013haa,Porto:2016pyg,Schafer:2018kuf,Levi:2018nxp}. Here we follow Ref.~\cite{Kol:2007bc}.
\footnote{Following the ideas in \cite{Goldberger:2004jt}.} A series of technical 
details for the calculation of the potential terms have already been given in 
Refs.~\cite{Blumlein:2019zku,Blumlein:2020pog} before. In 
the case of the tail terms one first applies the multi--pole expansion valid for the far zone  
\cite{Thorne:1980ru,Blanchet:1985sp,Blanchet:1987wq,
Blanchet:1993ec,Goldberger:2005cd,Ross:2012fc,Damour:2014jta,Galley:2015kus,Blanchet:2013haa,Schafer:2018kuf,Marchand:2020fpt,
Bini:2020wpo,Bini:2020nsb,Almeida:2020mrg} to the respective post--Newtonian order and then applies EFT methods to 
calculate their contribution, cf.~\cite{Foffa:2019eeb}. Expansions of this type generally belong to the operator product
expansions \cite{Wilson:1969zs}. In the calculation one also applies the method of expansion 
by regions \cite{Beneke:1997zp,Jantzen:2011nz}. 

In the present paper observables at 5PN such as the energy and periastron 
advance at circular orbits could not yet be calculated in complete form, since a series of differences with the
literature have still to be fully clarified. This concerns rational terms contributing to the tail term. However,
we obtain all other contributions, including the $\pi^2$ contributions to the yet undetermined constants $\bar{d}_5$ and 
$a_6$, in \cite{Bini:2020wpo}. Furthermore, quite a series of comparisons could be performed with the literature.

The paper is organized as follows. In Section~\ref{sec:2} we describe the calculation of the 5PN potential 
terms and present the associated Hamiltonian $H_{\rm 5PN}^{\rm pot}$ in the harmonic gauge. We use dimensional 
regularization  in $D = 4 - 2\ep$ dimensions. It is this method which allows a particular elegant merging of the 
potential and tail contributions in the conservative Hamiltonian, as we will show below. Already at 3PN the contributions 
to $H^{\rm pot}$ have poles in $1/\ep$, cf.~\cite{Foffa:2011ub}. From 4PN corresponding poles also appear in the tail 
terms. We will discuss the main aspects of the 5PN tail term in Section~\ref{sec:3} and construct a pole--free 
Hamiltonian 
in Section~\ref{sec:4}. Here we show that the poles in the combined Hamiltonian can be transformed away by a canonical 
transformation. In Section~\ref{sec:5} we compare to results given in the literature and discuss open questions.
A canonical transformation from harmonic to EOB coordinates is performed. We derive the non--local tail contributions
within our approach and calculate their contribution to the binding energy and to periastron advance in the circular 
case. We then turn to the contributions to periastron advance from the potential terms and derive the $\pi^2$ 
contributions to the previously unknown constants $\bar{d}_5$ and $a_6$ and summarize our present results for 
the circular binding energy and periastron advance. Furthermore we briefly discuss the remaining contributions to the 
tail term.
Section~\ref{sec:6} contains the conclusions. In the appendices some technical aspects are presented on the merging of 
the potential and tail terms using the method of expansion  by regions. As well we present longer formulae, which 
are used in the present calculation.
%----------------------------------------------------------------------------------------------------------------
\section{The potential contributions to the Hamiltonian}
\label{sec:2}
%----------------------------------------------------------------------------------------------------------------

\vspace*{1mm}
\noindent
The calculation of the 5PN corrections is performed in the same way as has been described in 
Refs.~\cite{Blumlein:2019zku,Blumlein:2020pog}. Starting from the Einstein--Hilbert Lagrangian, we parameterize the 
metric $g_{\mu\nu}$ according to Ref.~\cite{Kol:2007bc}
in 
terms of scalar, vector and tensor fields, and work in the harmonic gauge.\footnote{At a later stage technical
steps will require to move away from the harmonic gauge, which we will explain in detail.} The 
Feynman diagrams are generated using {\tt QGRAF} \cite{Nogueira:1991ex}. The Lorentz algebra is carried out using 
{\tt Form} \cite{FORM} and we perform the integration by parts (IBP) reduction to master integrals using the code 
{\tt Crusher} \cite{CRUSHER}. Table~\ref{TAB1} gives an overview on the present calculation.
%--------------------------------------------------------------------------------------------------------
\begin{table}[H]\centering
%--------------------------------------------------------------------------------------------------------
\begin{tabular}{rrrrrr}
\toprule
\multicolumn{1}{c}{\#loops}            &
\multicolumn{1}{c}{QGRAF}        &
\multicolumn{1}{c}{source irred.}           &
\multicolumn{1}{c}{no source loops}  &
\multicolumn{1}{c}{no tadpoles} &
\multicolumn{1}{c}{masters} \\
\midrule
  0  &       3 &      3 &      3 &      3 & 0\\
  1  &      72 &     72 &     72 &     72 & 1\\
  2  &    3286 &   3286 &   3286 &   2702 & 1\\
  3  &   81526 &  62246 &  60998 &  41676 & 1\\
  4  &  545812 & 264354 & 234934 & 116498 & 7\\
  5  &  332020 & 128080 & 101570 &  27582 & 4\\
\bottomrule
\end{tabular}
%--------------------------------------------------------------------------------------------------------
\caption[]{\sf Numbers of contributing diagrams at the different loop levels and master integrals.} 
\label{TAB1}
\end{table}
%--------------------------------------------------------------------------------------------------------

\noindent
From the graphs 
generated by {\tt QGRAF} one has to remove the source reducible graphs, graphs with source loops and tadpoles.
In this way the 962719 initial diagrams reduce to 188533 diagrams. The computation time amounts to about one week, 
including the time for the IBP reduction, on an {\tt Intel(R) Xeon(R) CPU E5-2643 v4} and it grows exponentially with 
the loop order. Most of the CPU time is needed to perform the time derivatives. Only one non--trivial master integral 
contributes, see \cite{Foffa:2016rgu,Blumlein:2019zku}.

One first obtains a Lagrange function of $m$th order still  containing the accelerations $a_i$ and time derivatives 
thereof. They are removed by using first double zero insertions \cite{Schafer:1984mr,Damour:1985mt} together with partial 
integration and the remaining linear accelerations by a shift 
\cite{DW,Damour:1985mt,Damour:1990jh,Blanchet:2013haa}, cf.~\cite{Blumlein:2020pog}. 
By this operation we leave harmonic coordinates. A Legendre transformation leads then to the 
potential contributions of the  Hamiltonian, which still contains pole terms in the dimensional parameter $\ep$.\footnote{
As also the case in renormalizable quantum field theories, Langrangians and Hamiltonians are in general no observables and
are generally singular.} The reduced Hamiltonian in the cms is given by
%----------------------------------------------------------------------------------------------------------------
\begin{eqnarray}
\label{eq:HatH}
\hat{H} = \frac{H-M c^2}{\mu c^2},
\end{eqnarray}
%----------------------------------------------------------------------------------------------------------------
with $c$ the velocity of light, $M = m_1+m_2$ the rest mass of the binary system and $\mu = m_1 m_2/M$, 
%----------------------------------------------------------------------------------------------------------------
\begin{eqnarray}
\hat{H}_{\rm 5PN}^{\rm pot} \hspace*{-3mm} &=& \hspace*{-3mm}
-\frac{21 p^{12}}{1024}
+\frac{5}{16 r^6}
-\frac{125 p^2}{16 r^5}
-\frac{499 p^4}{64 r^4}
-\frac{161 p^6}{32 r^3}
-\frac{445 p^8}{256 r^2}
-\frac{77 p^{10}}{256 r}
+\frac{17 (p.n)^2}{4 r^5}
+\frac{29 p^2 (p.n)^2}{8 r^4}
\nonumber\\ &&
+\frac{21 p^4 (p.n)^2}{16 r^3}
+\frac{5 p^6 (p.n)^2}{32 r^2}
-\frac{(p.n)^4}{8 r^4}
+\frac{1}{\ep} \Biggl\{
        \nu ^2 \Biggl[
                -\frac{520909}{37800 r^6}
     -\frac{698242 p^2}{4725 r^5}
                +\frac{592957 p^4}{2520 r^4}
\nonumber\\ &&
 -\frac{13583 p^6}{336 r^3}
                +\frac{23569 (p.n)^2}{540 r^5}
                -\frac{1895597 p^2 (p.n)^2}{2520 r^4}
  +\frac{23047 p^4 (p.n)^2}{112 r^3}
                +\frac{16223 (p.n)^4}{28 r^4}
\nonumber\\ &&           
     -\frac{130 p^2 (p.n)^4}{r^3}
                -\frac{91 (p.n)^6}{6 r^3}
        \Biggr]
        +\nu  \Biggl[
                -\frac{272309}{12600 r^6}
                +\frac{22439 p^2}{12600 r^5}
	 -\frac{49023 p^4}{560 r^4}
                +\frac{1173 p^6}{80 r^3}
\nonumber\\ &&               
                -\frac{210947 (p.n)^2}{2520 r^5}
                +\frac{25169 p^2 (p.n)^2}{105 r^4}
                -\frac{2271 p^4 (p.n)^2}{80 r^3}
                -\frac{13059 (p.n)^4}{70 r^4}
    -\frac{81 p^2 (p.n)^4}{r^3}
\nonumber\\ &&            
                +\frac{77 (p.n)^6}{r^3}
        \Biggr]
        +\nu ^3 \Biggl[
                \frac{28811 p^2}{210 r^5}
                -\frac{297509 p^4}{2520 r^4}
                -\frac{6889 p^6}{360 r^3}
                -\frac{3068 (p.n)^2}{7 r^5}
  +\frac{352834 p^2 (p.n)^2}{315 r^4}
\nonumber\\ &&              
                +\frac{16538 p^4 (p.n)^2}{105 r^3}
                -\frac{304669 (p.n)^4}{240 r^4}
                -
                \frac{18979 p^2 (p.n)^4}{56 r^3}
                +\frac{2891 (p.n)^6}{12 r^3}
        \Biggr]
\Biggr\}
\nonumber\\ &&
+\nu  \Biggl[
        \frac{231 p^{12}}{1024}
        -\frac{253555919}{529200 r^6}
        -\frac{1457872519 p^2}{2116800 r^5}
        +\frac{2128837091 p^4}{1411200 r^4}
        +\frac{11206267 p^6}{141120 r^3}
        +\frac{937 p^8}{32 r^2}
\nonumber\\ &&      
  +\frac{805 p^{10}}{256 r}
        +\pi ^2 \Biggl(
                \frac{70399}{1152 r^6}
                +\frac{65291 p^2}{1152 r^5}
                -\frac{1328147 p^4}{12288 r^4}
                -\frac{7719 p^6}{4096 r^3}
                +\frac{6649 (p.n)^2}{576 r^5}
\nonumber\\ &&               
 +\frac{5042575 p^2 (p.n)^2}{6144 r^4}
                +\frac{58887 p^4 (p.n)^2}{4096 r^3}
                -\frac{3293913 (p.n)^4}{4096 r^4}
                -\frac{89625 p^2 (p.n)^4}{4096 r^3}
\nonumber\\ &&
                +\frac{42105 (p.n)^6}{4096 r^3}
        \Biggr)
        +\ln\left(\frac{r}{r_0}\right) \Biggl(
                -\frac{272309}{1050 r^6}
                +\frac{22439 p^2}{1260 r^5}
                -\frac{49023 p^4}{70 r^4}
                +\frac{3519 p^6}{40 r^3}
                -\frac{210947 (p.n)^2}{252 r^5}
\nonumber\\ &&          
                +\frac{201352 p^2 (p.n)^2}{105 r^4}
      -\frac{6813 p^4 (p.n)^2}{40 r^3}
                -\frac{52236 (p.n)^4}{35 r^4}
                -\frac{486 p^2 (p.n)^4}{r^3}
                +\frac{462 (p.n)^6}{r^3}
        \Biggr)
\nonumber\\ &&   
        +\frac{467022407 (p.n)^2}{2116800 r^5}
     -\frac{2385014243 p^2 (p.n)^2}{282240 r^4}
        -
        \frac{162949463 p^4 (p.n)^2}{235200 r^3}
        -\frac{589 p^6 (p.n)^2}{16 r^2}
\nonumber\\ &&   
        -\frac{35 p^8 (p.n)^2}{256 r}
     +\frac{1895797259 (p.n)^4}{235200 r^4}
        +\frac{31715507 p^2 (p.n)^4}{23520 r^3}
        +\frac{8951 p^4 (p.n)^4}{384 r^2}
\nonumber\\ &&   
        -\frac{627281 (p.n)^6}{960 r^3}
     -\frac{5117 p^2 (p.n)^6}{320 r^2}
        +\frac{159 (p.n)^8}{28 r^2}
\Biggr]
+\nu ^2 \Biggl[
        -\frac{231 p^{12}}{256}
        +\frac{295859}{1050 r^6}
\nonumber\\ &&   
        +\frac{1652383903 p^2}{529200 r^5}
        -\frac{420686323 p^4}{132300 r^4}
     +\frac{3605263 p^6}{29400 r^3}
        -\frac{11535 p^8}{128 r^2}
        -\frac{2865 p^{10}}{256 r}
\nonumber\\ &&      
  + \ln\left(\frac{r}{r_0}\right) \Biggl(
                -\frac{520909}{3150 r^6}
                -\frac{1396484 p^2}{945 r^5}
                +\frac{592957 p^4}{315 r^4}
          -\frac{13583 p^6}{56 r^3}
                +\frac{23569 (p.n)^2}{54 r^5}
\nonumber\\ &&              
                -\frac{1895597 p^2 (p.n)^2}{315 r^4}
                +\frac{69141 p^4 (p.n)^2}{56 r^3}
                +\frac{32446 (p.n)^4}{7 r^4}
  -\frac{780 p^2 (p.n)^4}{r^3}
                -\frac{91 (p.n)^6}{r^3}
        \Biggr)
\nonumber\\ &&           
        +\pi ^2 \Biggl(
                \frac{11573}{768 r^6}
                -\frac{121315 p^2}{768 r^5}
                +\frac{2076041 p^4}{12288 r^4}
                +\frac{29987 p^6}{4096 r^3}
     +\frac{200359 (p.n)^2}{768 r^5}
\nonumber\\ &&            
                -\frac{5962205 p^2 (p.n)^2}{6144 r^4}
                -
                \frac{172311 p^4 (p.n)^2}{4096 r^3}
                +\frac{2617363 (p.n)^4}{4096 r^4}
    +\frac{127125 p^2 (p.n)^4}{4096 r^3}
\nonumber\\ &&   
                +\frac{14175 (p.n)^6}{4096 r^3}
        \Biggr)
        -\frac{944072707 (p.n)^2}{264600 r^5}
        +\frac{35606467999 p^2 (p.n)^2}{2116800 r^4}
     -\frac{1945067 p^4 (p.n)^2}{2450 r^3}
\nonumber\\ &&      
        +\frac{4969 p^6 (p.n)^2}{64 r^2}
        +\frac{275 p^8 (p.n)^2}{256 r}
        -\frac{1742633989 (p.n)^4}{117600 r^4}
  +\frac{848889 p^2 (p.n)^4}{1568 r^3}
        +\frac{925 p^4 (p.n)^4}{24 r^2}
\nonumber\\ &&      
        +\frac{15 p^6 (p.n)^4}{128 r}
        +\frac{18031 (p.n)^6}{3360 r^3}
        -\frac{8331 p^2 (p.n)^6}{160 r^2}
  +\frac{751 (p.n)^8}{28 r^2}
\Biggr]
+\nu ^3 \Biggl[
        \frac{1617 p^{12}}{1024}
\nonumber\\ &&  
        -\frac{298537367 p^2}{151200 r^5}
        +\frac{617770201 p^4}{423360 r^4}
        +\frac{108551131 p^6}{4233600 r^3}
        +\frac{16283 p^8}{256 r^2}
      +\frac{3995 p^{10}}{256 r}
        +\pi ^2 
\nonumber\\ &&     
\times \Biggl(
                -\frac{2339 p^2}{192 r^5}
                +\frac{98447 p^4}{3072 r^4}
                -\frac{20259 p^6}{1024 r^3}
                -\frac{16111 (p.n)^2}{192 r^5}
                +\frac{131231 p^2 (p.n)^2}{1536 r^4}
\nonumber\\ &&
           +\frac{106947 p^4 (p.n)^2}{1024 r^3}
                -\frac{361499 (p.n)^4}{1024 r^4}
                -\frac{30075 p^2 (p.n)^4}{1024 r^3}
                -
                \frac{65625 (p.n)^6}{1024 r^3}
        \Biggr)
\nonumber\\ &&           
        +\ln\left(\frac{r}{r_0}\right) \Biggl(
                \frac{28811 p^2}{21 r^5}
                -\frac{297509 p^4}{315 r^4}
                -\frac{6889 p^6}{60 r^3}
                -\frac{30680 (p.n)^2}{7 r^5}
                +\frac{2822672 p^2 (p.n)^2}{315 r^4}
\nonumber\\ &&   
                +\frac{33076 p^4 (p.n)^2}{35 r^3}
     -\frac{304669 (p.n)^4}{30 r^4}
                -\frac{56937 p^2 (p.n)^4}{28 r^3}
                +\frac{2891 (p.n)^6}{2 r^3}
        \Biggr)
        +\frac{966353501 (p.n)^2}{151200 r^5}
\nonumber\\ &&   
     -\frac{3656476457 p^2 (p.n)^2}{235200 r^4}
        -\frac{2369976949 p^4 (p.n)^2}{1411200 r^3}
        +\frac{177 p^6 (p.n)^2}{256 r^2}
        -\frac{221 p^8 (p.n)^2}{64 r}
\nonumber\\ &&   
     +\frac{14035555739 (p.n)^4}{705600 r^4}
        +\frac{373945981 p^2 (p.n)^4}{94080 r^3}
        -\frac{125225 p^4 (p.n)^4}{768 r^2}
        -\frac{3 p^6 (p.n)^4}{128 r}
\nonumber\\ &&
     -\frac{14830647 (p.n)^6}{4480 r^3}
        +\frac{136977 p^2 (p.n)^6}{1280 r^2}
        -\frac{15 p^4 (p.n)^6}{128 r}
        -\frac{289839 (p.n)^8}{4480 r^2}
        -\frac{35 p^2 (p.n)^8}{256 r}
\Biggr]
\nonumber\\ &&    
+\nu ^4 \Biggl[
        -\frac{1155 p^{12}}{1024}
        -\frac{593 p^6}{32 r^3}
        +\frac{6649 p^8}{256 r^2}
        -\frac{1615 p^{10}}{256 r}
        +\frac{549 p^4 (p.n)^2}{32 r^3}
        -\frac{62143 p^6 (p.n)^2}{256 r^2}
\nonumber\\ &&    
    +\frac{867 p^8 (p.n)^2}{256 r}
        -\frac{5749 p^2 (p.n)^4}{96 r^3}
	        -\frac{3 p^6 (p.n)^4}{64 r}
+ \frac{652381 p^4 (p.n)^4}{768}         
-\frac{17623 (p.n)^6}{240 r^3}
\nonumber\\ &&
    -\frac{1178329 p^2 (p.n)^6}{1280 r^2}
        -\frac{45 p^4 (p.n)^6}{128 r}
        +\frac{1443091 (p.n)^8}{4480 r^2}
        +\frac{105 p^2 (p.n)^8}{128 r}
\Biggr]
+\nu ^5 \Biggl[
        \frac{231 p^{12}}{1024}
        -\frac{63 p^{10}}{256 r}
\nonumber\\ &&
        -\frac{35 p^8 (p.n)^2}{256 r}
        -\frac{15 p^6 (p.n)^4}{128 r}
        -\frac{15 p^4 (p.n)^6}{128 r}
        -\frac{35 p^2 (p.n)^8}{256 r}
        -\frac{63 (p.n)^{10}}{256 r}
\Biggr],
\end{eqnarray}
%----------------------------------------------------------------------------------------------------------------
with
%----------------------------------------------------------------------------------------------------------------
\begin{eqnarray}
r_0 = \frac{e^{-\gamma_E/2}}{2\sqrt{\pi} \mu_1},
\end{eqnarray}
%----------------------------------------------------------------------------------------------------------------
where $\gamma_E$ is the Euler--Mascheroni constant and $\mu_1$ the mass scale accounting for Newton's 
constant $G_N \rightarrow G_N \mu_1^{-2\ep}$ in $D$ dimensions. The corresponding contributions up to 4PN 
have been presented in \cite{Blumlein:2020pog} before. We rescale 
%----------------------------------------------------------------------------------------------------------------
\begin{eqnarray}
p = p_{\rm phys}/(\mu c),~~~
r = \left(G_N M/ c^2\right) r_{\rm phys},
\end{eqnarray}
%----------------------------------------------------------------------------------------------------------------
where $p$ and $r$ are now the rescaled (dimensionless) cms momentum and the distance of the two masses,
with $\vec{n} = \vec{r}/r$. In the following we will as widely as possible work with dimensionless quantities.

Pole and logarithmic contributions appear at $O(\nu), O(\nu^2)$ and $O(\nu^3)$, in accordance with the lower 
PN orders, where also always one more order in $\nu$ contributes from 3PN onward. We will see in Section~\ref{sec:4} 
that  the tail term is only singular for $O(\nu)$ and  $O(\nu^2)$ at 5PN.

In the Schwarzschild limit, $\nu \rightarrow 0$, one obtains the following contributions, 
%----------------------------------------------------------------------------------------------------------------
\begin{eqnarray}
\hat{H}^{\rm Schw}_{\rm 5PN} &=& 
        -\frac{21 p^{12}}{1024}
        -\frac{77 p^{10}}{256 r}
        -\frac{445 p^8}{256 r^2}
        -\frac{161 p^6}{32 r^3}
        -\frac{499 p^4}{64 r^4}
        -\frac{125 p^2}{16 r^5}
        +\frac{5 p^6 (p.n)^2}{32 r^2}
        +\frac{21 p^4 (p.n)^2}{16 r^3}
\nonumber\\ &&      
  +\frac{29 p^2 (p.n)^2}{8 r^4}
        -\frac{(p.n)^4}{8 r^4}
        +\frac{17 (p.n)^2}{4 r^5}
        +\frac{5}{16 r^6},
\end{eqnarray}
%----------------------------------------------------------------------------------------------------------------
in agreement with the expansion of Eq.~(30), \cite{Blumlein:2020pog}, to 5PN, cf.~\cite{SCHMUTZER1,WEINB}.
%----------------------------------------------------------------------------------------------------------------
\section{Remarks on the tail term}
\label{sec:3}
%----------------------------------------------------------------------------------------------------------------

\vspace*{1mm}
\noindent
We will derive a pole--free Hamiltonian at 5PN in Section~\ref{sec:4}. For this we will add the singular and 
logarithmic terms of the tail term, $\hat{H}_{\rm 5PN}^{\rm tail, sing, log}$, to the potential term $\hat{H}_{\rm 
5PN}^{\rm pot}$. Since these contributions are calculated, by different methods, either in the far zone (FZ) or 
the near zone (NZ), the question arises whether potential overlap contributions have to be considered. 
We remind that the calculation is performed in $D$ dimensions, not using any other regularization.

One may apply the method of expansion by regions, which has been introduced for the asymptotic expansion 
of Feynman integrals 
for bound states in the non--relativistic limit in~\cite{Beneke:1997zp,Jantzen:2011nz}. Here each loop integral 
is split into four {\it distinct} momentum regions, which are denoted as hard, soft, potential, and ultrasoft. 
Integrals over the hard and soft region correspond to quantum corrections and are not considered in the context 
of classical gravity.

The potential region, characterized by the momentum scaling 
%------------------------------------------------------------------------------------------------------------------------
\begin{equation}
|k_0| \sim \frac{v}{R},~~~~~~|k_i| \sim \frac{1}{R}~,
\end{equation}
%------------------------------------------------------------------------------------------------------------------------
with $v \in [v_1, v_2]$ and $v = v_{\rm phys}/c$ the typical velocities, is also referred to as orbital region. 
Here $k_i$ and $R$ are not rescaled. However, we set the associated action variable to 1.\footnote{In the quantum field 
theoretic case this would correspond to $\hbar = 1$.}
It can be identified 
with the near zone of the literature (i.e. the potential terms). In the ultrasoft (or radiation region), 
corresponding to the far zone (i.e. the tail terms),  momenta exhibit the uniform four--momentum scaling 
%------------------------------------------------------------------------------------------------------------------------
\begin{equation}
|k_\mu| \sim \frac{v}{R}~. 
\end{equation} 
%------------------------------------------------------------------------------------------------------------------------ 
The kinematic region of the potential term is 
%------------------------------------------------------------------------------------------------------------------------
\begin{equation}
\label{eq:POT}
k_i \in D_{\rm pot} =\left[-\infty,-\tfrac{1}{R}\right] \cup \left[\tfrac{1}{R},\infty\right], 
\end{equation} 
%------------------------------------------------------------------------------------------------------------------------ 
with $R$ of the order of separation the binary system. Likewise, the one of the tail term is 
%------------------------------------------------------------------------------------------------------------------------
\begin{equation}
\label{eq:US}
k_i \in D_{\rm us} = \left[-\tfrac{1}{R},\tfrac{1}{R}\right]. 
\end{equation} 
%------------------------------------------------------------------------------------------------------------------------ 
In the former region the exchanged fields are potential gravitons and in the latter region ultrasoft gravitons. 
One performs a Taylor expansion of the integrands according to the respective momentum scaling in $v$ up to the 
respective post--Newtonian order by observing that 
%------------------------------------------------------------------------------------------------------------------------ 
\begin{equation} 
v^2 \sim \frac{1}{r}.
\end{equation} 
%------------------------------------------------------------------------------------------------------------------------ 
For the tail terms this expansion includes the multi--pole expansion, which we will discuss below.\footnote{
As is often the case in EFT representations, the corresponding expansions are not just kinematic. An important example
in this respect is the light-cone expansion \cite{LCE}. In the most simple case of the twist--2 contributions its 
results are also obtained by the QCD improved parton model, resulting from a kinematic expansion. This is much more
subtle at higher twist, where partonic pictures require further conditions to give the same result, cf. 
\cite{Blumlein:2012bf} for a survey.}
Let us introduce the  operators $T_{\rm pot}$ and $T_{\rm us}$, which describe the Taylor 
expansions (with a few Laurent--terms in some cases) in the potential region and the ultrasoft region. In the 
post--Newtonian expansion they are given, more precisely, by
%--------------------------------------------------------------------------------------------------------- 
\begin{eqnarray} 
\label{eq:TN}
T^{\rm N}_i I_0(v) := \theta(N) \sum_{k=0}^\infty T_{i,k} v^k,
\end{eqnarray} 
%------------------------------------------------------------------------------------------------------------------------ 
with the quantifier $\theta(N)$ truncating the series at a maximal term $v^N, N \in \mathbb{N}$, which is idempotent
$\theta^l(N) \equiv \theta(N)$. Here the coefficients $T_{i,k}$ denote the expansion coefficients of the function
$I_0(v)$. 
The corresponding integrals have the following form 
%--------------------------------------------------------------------------------------------------------- 
\begin{eqnarray} 
\label{eq:I1}
I_1 = \int_{D_{\rm pot}} d k_i T_{\rm pot}^N I + \int_{D_{\rm us}} d k_i T_{\rm us}^N I
\end{eqnarray} 
%------------------------------------------------------------------------------------------------------------------------ 
for each of the $D$ components $k_i$,
where $I$ denotes the original integrand. One further obtains
%--------------------------------------------------------------------------------------------------------- 
\begin{eqnarray} 
I_1 = 
\int_{-\infty}^{+\infty} d k_i T_{\rm pot}^N I 
- \int_{D_{\rm us}} d k_i T_{\rm pot}^N I
+\int_{-\infty}^{+\infty} d k_i T_{\rm us}^N I 
- \int_{D_{\rm pot}} d k_i T_{\rm us}^N I. 
\end{eqnarray} 
%------------------------------------------------------------------------------------------------------------------------ 
In the respective domains $D_{\rm pot}$ and $D_{\rm us}$ one may further apply the operators
$T_{\rm pot}$ and $T_{\rm us}$ given the post--Newtonian accuracy one is working in. One then obtains
%--------------------------------------------------------------------------------------------------------- 
\begin{eqnarray} 
I_1 &=& 
\int_{-\infty}^{+\infty} d k_i T_{\rm pot}^N I 
- \int_{D_{\rm us}} d k_i T_{\rm us} T_{\rm pot}^N I
+\int_{-\infty}^{+\infty} d k_i T_{\rm us}^N I 
- \int_{D_{\rm pot}} d k_i T_{\rm pot} T_{\rm us}^N I. 
\label{eq:I1b}
\end{eqnarray} 
%------------------------------------------------------------------------------------------------------------------------ 
Here the 2nd and 4th term are the overlap integrals. Eq.~(\ref{eq:I1b}) can be further arranged to
%--------------------------------------------------------------------------------------------------------- 
\begin{eqnarray} 
I_1 &=&
\int_{-\infty}^{+\infty} d k_i T_{\rm pot}^N I 
+\int_{-\infty}^{+\infty} d k_i T_{\rm us}^N I 
- \int_{-\infty}^{+\infty} d k_i T_{\rm pot}^N T_{\rm us}^N I, 
\label{eq:I1a}
\end{eqnarray} 
%------------------------------------------------------------------------------------------------------------------------ 
provided that 
%------------------------------------------------------------------------------------------------------------------------ 
\begin{eqnarray} 
  T_{\rm us}^N T_{\rm pot}^N - T_{\rm pot}^N T_{\rm us}^N = 0
\end{eqnarray} 
%------------------------------------------------------------------------------------------------------------------------ 
holds,
which we prove in Appendix~\ref{sec:A}. Furthermore, the operation $T_{\rm us} T_{\rm pot}$ leads to scaleless 
integrands, implying that the last term in Eq.~(\ref{eq:I1a}) vanishes in $D$ dimensions, see Appendix~\ref{sec:A}.

We finally would would like to make some remarks on the relation on the multi--pole expansion \cite{Goldberger:2004jt} in 
the far zone to the ultrasoft region. One is starting from the full theory of general relativity in harmonic coordinates, 
i.e. the bulk action
%------------------------------------------------------------------------------------------------------------------------ 
\begin{equation}
  \label{eq:S_GR}
  S_{\text{GR},\text{bulk}} = 2\Lambda^2 \int d^Dx\ \sqrt{-g}\left(R - \frac{1}{2} \Gamma^\mu \Gamma_\mu\right).
\end{equation}
%------------------------------------------------------------------------------------------------------------------------ 
Here $\Lambda= c^2 \mu_1^\ep/\sqrt{32 \pi G_N}$, $R$ is the Ricci scalar,  $\Gamma^\mu = \Gamma^\mu_{\alpha\beta} 
g^{\alpha\beta}$ with 
$\Gamma^\mu_{\alpha\beta} = \tfrac{1}{2} g^{\mu\gamma} (g_{\gamma\alpha,\beta} +g_{\gamma\beta,\alpha} 
- g_{\alpha\beta,\gamma})$ is the Christoffel symbol. (\ref{eq:S_GR}) is coupled to compact objects via the action
%------------------------------------------------------------------------------------------------------------------------ 
\begin{equation}
  \label{eq:S_pp}
  S_{\text{pp}} = -\sum_{a=1}^2 m_a \int d\tau_a,
\end{equation}
%------------------------------------------------------------------------------------------------------------------------ 
with proper times $\tau_1, \tau_2$, one decomposes the metric into
%------------------------------------------------------------------------------------------------------------------------ 
\begin{equation}
  \label{eq:g_dec}
  g_{\mu\nu} = \eta_{\mu\nu} + \frac{1}{\Lambda}(H_{\mu\nu} + h_{\mu\nu})\,,
\end{equation}
%------------------------------------------------------------------------------------------------------------------------ 
where $\eta_{\mu\nu}$ is the Minkowski metric. The momenta associated with $H_{\mu\nu}$ are of the potential type and the 
momenta of the $h_{\mu\nu}$ fields are ultrasoft, cf. Eqs.~(\ref{eq:POT}) and (\ref{eq:US}). The resulting loop 
integrals therefore have the same form as Eq.~(\ref{eq:I1}), as obtained from the asymptotic expansion.

The action of the full theory is matched to the non--relativistic general relativity (NRGR) action by
%------------------------------------------------------------------------------------------------------------------------ 
\begin{align}
  \label{eq:S_NRGR}
  S_{\text{NRGR}} ={}&S_{\text{NRGR},\text{bulk}} + S_{\text{NRGR}, h^0} 
+ S_{\text{NRGR}, h^1} + \mathcal{O}\left(h^2\right)\,,
\end{align}
%------------------------------------------------------------------------------------------------------------------------ 
with
%------------------------------------------------------------------------------------------------------------------------ 
\begin{align}
  \label{eq:S_NRGR_h0}
  S_{\text{NRGR}, h^0} ={}& \int dt\ (T - V_{\text{NZ}})\,, \\
  \label{eq:S_NRGR_h1}
  S_{\text{NRGR}, h^1} ={}& \frac{1}{2\Lambda} \int d^D x\ T^{\mu\nu}h_{\mu\nu}\,,
\end{align}
%------------------------------------------------------------------------------------------------------------------------ 
where there are no potential modes anymore. Here $T$ denotes the kinetic term and  $V_{\text{NZ}}$ the near--zone potential. 
$S_{\text{NRGR},\text{bulk}}$ is the same as the general relativity bulk action $S_{\text{GR},\text{bulk}}$ from 
Eq.~\eqref{eq:S_GR}, but without the potential contributions to the metric. Both $V_{\text{NZ}}$ and the effective 
stress--energy 
tensor $T^{\mu\nu}$ are fixed by requiring that the NRGR action produces the same predictions as the asymptotic expansion of the 
full theory. In other words, the integrals over the potential region are absorbed into $V_{\text{NZ}}$ and
$T^{\mu\nu}$.

We now elaborate on the relation to the multi--pole expansion. Consider
%------------------------------------------------------------------------------------------------------------------------ 
\begin{equation}
  \label{eq:h_fourier}
  h_{\mu\nu}(x) = \int \frac{d^D p}{(2\pi)^D} e^{ipx} h_{\mu\nu}(p)\,,
\end{equation}
%------------------------------------------------------------------------------------------------------------------------ 
where the momentum is ultrasoft by definition, i.e. 
%------------------------------------------------------------------------------------------------------------------------ 
\begin{equation}
  \label{eq:px_scaling}
  \vec{p}\,\vec{x} \sim v.
\end{equation}
%------------------------------------------------------------------------------------------------------------------------ 
We can Taylor expand the exponential in Eq.~\eqref{eq:h_fourier} to obtain
%------------------------------------------------------------------------------------------------------------------------ 
\begin{equation}
  \label{eq:h_us_exp}
  h_{\mu\nu}(x) = \int \frac{d^Dp}{(2\pi)^D} e^{-i p_0 x_0} \left( \sum_{n=0}^N \frac{(i\vec{p}\vec{x})^n}{n!} + 
\mathcal{O}\left(v^{N+1}\right)\right)h_{\mu\nu}(p)\,.
\end{equation}
%------------------------------------------------------------------------------------------------------------------------ 
Rewriting
%------------------------------------------------------------------------------------------------------------------------ 
\begin{equation}
  \label{eq:p_to_dx}
  i^n p_{i_1} \dots p_{i_n} = \Big[\partial x_{i_1} \dots \partial x_{i_n} e^{i\vec{p}\vec{x}} \Big]_{\vec{x} = 0}
\end{equation}
%------------------------------------------------------------------------------------------------------------------------ 
now yields
%------------------------------------------------------------------------------------------------------------------------ 
\begin{align}
h_{\mu\nu}(x)={}&  \sum_{n=0}^N \frac{1}{n!} x_{i_1} \dots x_{i_n} 
\left[\partial x_{i_1} \dots \partial x_{i_n} \int_{-\infty}^\infty \frac{d^Dp}{(2\pi)^D} e^{ipx} h_{\mu\nu}(p) \right]_{\vec{x} 
= 0}+ \mathcal{O}\left(v^{N+1}\right)\notag\\
     ={}&  \sum_{n=0}^N \frac{1}{n!} x_{i_1} \dots x_{i_n} \Big[\partial x_{i_1} \dots \partial x_{i_n} h_{\mu\nu}(x) \Big]_{\vec{x} = 0}
+ \mathcal{O}\left(v^{N+1}\right).
\end{align}
%------------------------------------------------------------------------------------------------------------------------ 
Inserting this expression back into the linear ultrasoft action
Eq.~(\ref{eq:S_NRGR_h1}) we
retrieve the familiar starting point of the multi--pole expansion
%------------------------------------------------------------------------------------------------------------------------ 
\begin{equation}
  \label{eq:S_us1_multipole}
  S_{\text{NRGR}, h^1} = \frac{1}{2\Lambda} \int d^Dx\ T^{\mu\nu}(x) \sum_{n=0}^N \frac{1}{n!} x_{i_1} \dots x_{i_n} \Big[\partial x_{i_1} \dots \partial x_{i_n} h_{\mu\nu}(x) \Big]_{\vec{x} = 0}+ \mathcal{O}\left(v^{N+1}\right)\,,
\end{equation}
%------------------------------------------------------------------------------------------------------------------------ 
with an explicit velocity power counting.
The remaining steps in the multi--pole expansion are standard. In short,
one defines moments
%------------------------------------------------------------------------------------------------------------------------ 
\begin{equation}
  M_n^{\mu\nu} = \int d^{D-1}\vec{x}\ T^{\mu\nu}(x) x_{i_1}\dots x_{i_n}\,,
\end{equation}
%------------------------------------------------------------------------------------------------------------------------ 
and decomposes them into irreducible $SO(3)$ spherical tensors, choosing a
symmetric trace--free (STF) basis, e.g.~\cite{Ross:2012fc,Damour:2014jta}.
The tail terms are represented by the multi--pole expansion valid in the far zone. In our treatment we will follow 
Refs.~\cite{Foffa:2019eeb,Marchand:2020fpt}. 
%---------------------------------------------------------------------------------------------------------------- 
\section{The pole-free Hamiltonian at 5PN} \label{sec:4} 
%----------------------------------------------------------------------------------------------------------------

\vspace*{1mm}
\noindent
It is convenient to work with pole--free Hamiltonians and we add the singular and logarithmic pieces of the 
Hamiltonian of the tail term in 5PN, $\hat{H}_{\rm 5PN}^{\rm tail, sing, log}$, 
%----------------------------------------------------------------------------------------------------------------
\begin{eqnarray}
\hat{H}_{\rm 5PN}^{\rm tail, sing, log} \hspace*{-2mm} &=& 
\frac{1}{\ep} \Biggl\{
\Biggl(\frac{16 \nu}{105} - \frac{332 \nu^2}{105}\Biggr) \frac{1}{r^6}
+
\Biggl[\Biggl(\frac{236 \nu}{35} - \frac{212 \nu^2}{35}\Biggr)p^2 
- \Biggl(\frac{684 \nu}{35} + \frac{1264 \nu^2}{105}\Biggr) (p.n)^2\Biggr] \frac{1}{r^5}
\nonumber\\ &&
+ 
\Biggl[
  \Biggl(\frac{533 \nu}{21} + \frac{706 \nu^2}{21}\Biggr) p^4 
- \Biggl(\frac{7732\nu}{35} + \frac{10936 \nu^2}{105}\Biggr) p^2 (p.n)^2 
+ \Biggl(\frac{6197\nu}{35} + \frac{2656\nu^2}{35}\Biggr)
\times \nonumber\\ && 
(p.n)^4\Biggr] \frac{1}{r^4} 
+ \Biggl[\Biggl(\frac{94 \nu}{15} - \frac{94 \nu^2}{5}\Biggr) p^6 
+ \Biggl(-\frac{172 \nu}{5} + \frac{516 \nu^2}{5}\Biggr) p^4 (p.n)^2
+ \Biggl(26 \nu - 78 \nu^2\Biggr) 
\nonumber\\ && 
\times p^2 (p.n)^4 \Biggr] \frac{1}{r^3}\Biggr\}
\nonumber\\ && 
+ \Biggl\{
  \Biggl[
  \frac{128 \nu}{105} - \frac{2656 \nu^2}{105} \Biggr]\frac{1}{r^6} 
+ \Biggl[
  \Biggl(\frac{1416 \nu}{35} - \frac{1272 \nu^2}{35} \Biggr) p^2 
- \Biggl(\frac{4104\nu}{35} + \frac{2528 \nu^2}{35} \Biggr)
\nonumber\\ && \times
(p.n)^2 \Biggr] \frac{1}{r^5}
+ \Biggl[\Biggl(\frac{2132 \nu}{21} + \frac{2824 \nu^2}{21}\Biggr) p^4  
- \Biggl(\frac{30928 \nu}{35} + \frac{43744 \nu^2}{105}\Biggr) p^2 (p.n)^2 
\nonumber\\ &&
+ \Biggl(\frac{24788 \nu}{35} + \frac{10624 \nu^2}{35}\Biggr) (p.n)^4 \Biggr] \frac{1}{r^4}
+ \Biggl[\Biggl(\frac{188 \nu}{15} - \frac{188 \nu^2}{5}\Biggr) p^6 
+ \Biggl(-\frac{344 \nu}{5} 
\nonumber\\ && + \frac{1032 \nu^2}{5}\Biggr) 
p^4 (p.n)^2
+ \Biggl(52 \nu - 156 \nu^2\Biggr) p^2 (p.n)^4\Biggr] \frac{1}{r^3} \Biggr\} 
\ln\left(\frac{r}{r_0}\right)
\label{eq:Htail1}
\end{eqnarray}
%----------------------------------------------------------------------------------------------------------------
to $\hat{H}_{\rm 5PN}^{\rm pot}$. For all contributions resulting into Eq.~(\ref{eq:Htail1}) we agree with the
integrals in the multi--pole expansion in \cite{Foffa:2019eeb}.

The sum of the potential term and this contribution is not pole--free yet, as is the case from 3PN onward, cf. 
\cite{Blumlein:2020pog}. However, after performing the following canonical transformation a pole--free Hamiltonian
is obtained, which is not the case for $H_{\rm 5PN}^{\rm pot}$ and $H_{\rm 5PN}^{\rm tail, sing}$ individually.
By this transformation one further moves away from the harmonic coordinates, which were used at the starting point 
of the calculation. Still a prediction of all observables is possible.
Moreover, the comparison with EOB results becomes simpler, since they are given in pole--free form \cite{Bini:2020wpo}.

Following the formalism described in Ref.~\cite{Blumlein:2020pog}, Eqs. (38--41), one obtains
the corresponding generating function 
%----------------------------------------------------------------------------------------------------------------
\begin{eqnarray}
G(p^2,p.n,r;\ep) &=& p.n\Biggl\{\frac{1}{\ep} \Biggl\{
             - t_3 \frac{17 \nu}{6 r^2}
        + t_4 \Biggl[
                \frac{1}{r^2} \Biggl(\frac{1}{90} \nu  (585+4 \nu ) p^2 
                -\frac{1}{3} \nu  (12+37 \nu ) (p.n)^2\Biggr)
\nonumber\\ &&               
 -\frac{\nu  (65+264 \nu )}{30 r^3}
        \Biggr]
             + t_5 \Biggl[
                \frac{1}{r^3} \Biggl(
                        -\frac{1}{420} \nu  \big(
                                -43-69088 \nu +52078 \nu ^2\big) p^2 
\nonumber\\ &&                        
+\frac{\nu  \big(
                                1785-204868 \nu +264212 \nu ^2\big) (p.n)^2}{1260}
                \Biggr)
                +\frac{1}{r^2} \Biggl(
\nonumber\\ &&                     
   -\frac{\nu  \big(
                                -127218+374300 \nu +70007 \nu ^2\big) p^4}{5040}
-\frac{1}{12} \nu  \big(
                                132-26 \nu +413 \nu ^2\big) (p.n)^4
\nonumber\\ &&
                        +\frac{\nu  \big(
 -1680+36624 \nu +43567 \nu ^2\big) p^2 (p.n)^2}{840}
                \Biggr)
  +\frac{\nu  (889917+388114 \nu )}{37800 r^4}
        \Biggr]
\Biggr\}
\Biggr\}.
\nonumber\\ 
\end{eqnarray}
%----------------------------------------------------------------------------------------------------------------
%----------------------------------------------------------------------------------------------------------------

Furthermore, we transform the logarithmic part to explicitly match the structure of the non--local contribution 
from the tail term in harmonic coordinates, 
%----------------------------------------------------------------------------------------------------------------
 \begin{equation}
  \delta H^{\mathrm{4+5PN}}_\mathrm{log} = 2 \frac{G^3_N E}{c^{10}} \left(
\frac{1}{5} I^{(3)}(t)^2 +\frac{1}{189 c^2} O^{(4)}(t)^2 + \frac{16}{45
c^2} J^{(3)}(t)^2   \right)\ln\left( \frac{r}{r_0} \right)\, ,
\end{equation}
%----------------------------------------------------------------------------------------------------------------
see also Section~\ref{sec:52}. Here the multi--pole moments $I_{ab}, O_{abc}$ and $J_{ab}$ are those of Eq.~(2.4) in 
\cite{Bini:2020wpo}, with indices contracted, and $E$ is the total energy.  The corresponding transformation reads
%----------------------------------------------------------------------------------------------------------------
\begin{eqnarray}
G(p^2,p.n,r;\ln(r/r_0)) &=& p.n \Biggl\{\ln \left(\frac{r}{r_0}\right) \Biggl\{
        -t_3 \frac{17 \nu}{r^2}
        +t_4 \Biggl[
                \frac{1}{r^2} \Biggl(\frac{\nu  (585+4 \nu ) p^2}{15}
                -2 \nu  (12+37 \nu ) 
\nonumber\\ && \times
(p.n)^2
                \Biggr)
                -\frac{4 \nu  (73+264 \nu )}{15 r^3}
        \Biggr]
+
        t_5 \Biggl[
                \frac{1}{r^3} \Biggl(
                        -\frac{2}{105} \nu  \big(
                                7557-65496 \nu 
\nonumber\\ &&
+52078 \nu ^2\big) p^2 
                        +\frac{2}{315} \nu  \big(
                                21345-195484 \nu +264212 \nu ^2\big) (p.n)^2
                \Biggr)
\nonumber\\ &&
                +\frac{1}{r^2} \Biggl(
                        -\frac{1}{840} \nu  \big(
                                -106162+311132 \nu +70007 \nu ^2\big) p^4
                        +\frac{1}{140} \nu  \big(
                                1232
\nonumber\\ &&
+27888 \nu +43567 \nu ^2\big) p^2 (p.n)^2
                        -\frac{1}{2} \nu  \big(
                                132-26 \nu +413 \nu ^2\big) (p.n)^4
                \Biggr)
\nonumber\\ &&
                +\frac{\nu  (886461+331090 \nu )}{3780 r^4}
        \Biggr]
\Biggr\}
\Biggr\}.
\end{eqnarray}
%----------------------------------------------------------------------------------------------------------------
Here $t_i, i = 1 ... 5$ labels the $i$th post--Newtonian order.

The pole--free Hamiltonian based on the above contributions is then given by\footnote{Note that $\hat{H}_{\rm 5PN}^{\rm pole 
free}$ does not yet contain the complete local Hamiltonian.}
%----------------------------------------------------------------------------------------------------------------
\begin{eqnarray}
\hat{H}_{\rm 5PN}^{\rm pole free} &=&
-
\frac{21 p^{12}}{1024}
+\frac{5}{16 r^6}
-\frac{125 p^2}{16 r^5}
-\frac{499 p^4}{64 r^4}
-\frac{161 p^6}{32 r^3}
-\frac{445 p^8}{256 r^2}
-\frac{77 p^{10}}{256 r}
\nonumber\\ &&
+\nu  \Biggl[
        \frac{231 p^{12}}{1024}
        -\frac{279775133}{529200 r^6}
        -\frac{1450584679 p^2}{2116800 r^5}
        +\frac{2010713771 p^4}{1411200 r^4}
        +\frac{11206267 p^6}{141120 r^3}
        +\frac{937 p^8}{32 r^2}
\nonumber\\ &&     
    +\frac{805 p^{10}}{256 r}
        +\ln\left(\frac{r}{r_0}\right) \Biggl(
                \frac{64}{105 r^6}
                -\frac{18944 p^2}{105 r^5}
                +\frac{1796 p^4}{105 r^4}
                +\frac{19136 (p.n)^2}{105 r^5}
                -\frac{10664 p^2 (p.n)^2}{105 r^4}
\nonumber\\ &&
                +\frac{2748 (p.n)^4}{35 r^4}
        \Biggr)
        +\pi ^2 \Biggl(
                \frac{70399}{1152 r^6}
                +\frac{65291 p^2}{1152 r^5}
                -\frac{1328147 p^4}{12288 r^4}
                -\frac{7719 p^6}{4096 r^3}
                +\frac{6649 (p.n)^2}{576 r^5}
\nonumber\\ &&              
  +\frac{5042575 p^2 (p.n)^2}{6144 r^4}
                +\frac{58887 p^4 (p.n)^2}{4096 r^3}
                -\frac{3293913 (p.n)^4}{4096 r^4}
                -\frac{89625 p^2 (p.n)^4}{4096 r^3}
\nonumber\\ &&              
  +\frac{42105 (p.n)^6}{4096 r^3}
        \Biggr)
        -\frac{34541593 (p.n)^2}{2116800 r^5}
        -\frac{2395722563 p^2 (p.n)^2}{282240 r^4}
        -\frac{62196341 p^4 (p.n)^2}{78400 r^3}
\nonumber\\ &&     
   -
        \frac{589 p^6 (p.n)^2}{16 r^2}
        -\frac{35 p^8 (p.n)^2}{256 r}
        +\frac{631107353 (p.n)^4}{78400 r^4}
        +\frac{31226291 p^2 (p.n)^4}{23520 r^3}
\nonumber\\ &&      
  +\frac{8951 p^4 (p.n)^4}{384 r^2}
        -\frac{563921 (p.n)^6}{960 r^3}
        -\frac{5117 p^2 (p.n)^6}{320 r^2}
        +\frac{159 (p.n)^8}{28 r^2}
\Biggr]
+\nu ^2 \Biggl[
        -\frac{231 p^{12}}{256}
\nonumber\\ &&      
  +\frac{72454}{225 r^6}
        +\frac{1353196483 p^2}{529200 r^5}
        -\frac{787300061 p^4}{264600 r^4}
        +\frac{3605263 p^6}{29400 r^3}
        -\frac{11535 p^8}{128 r^2}
        -\frac{2865 p^{10}}{256 r}
\nonumber\\ && 
 - \ln\left(\frac{r}{r_0}\right) \Biggl(
                \frac{256}{105 r^6}
                +\frac{3392 p^2}{105 r^5}
                -\frac{432 p^4}{35 r^4}
                -\frac{2992 (p.n)^2}{105 r^5}
                -\frac{6824 p^2 (p.n)^2}{105 r^4}
                +\frac{496 (p.n)^4}{7 r^4}
        \Biggr)
\nonumber\\ &&
        +\pi ^2 \Biggl(
                \frac{5453}{768 r^6}
                -\frac{121315 p^2}{768 r^5}
                +\frac{2076041 p^4}{12288 r^4}
                +\frac{29987 p^6}{4096 r^3}
                +\frac{200359 (p.n)^2}{768 r^5}
                -\frac{172311 p^4 (p.n)^2}{4096 r^3}
\nonumber\\ &&              
  -\frac{5962205 p^2 (p.n)^2}{6144 r^4}
                +\frac{2617363 (p.n)^4}{4096 r^4}
                +\frac{127125 p^2 (p.n)^4}{4096 r^3}
                +\frac{14175 (p.n)^6}{4096 r^3}
        \Biggr)
\nonumber\\ &&       
 -\frac{857318207 (p.n)^2}{264600 r^5}
        +
        \frac{34200172759 p^2 (p.n)^2}{2116800 r^4}
        -\frac{5034763 p^4 (p.n)^2}{9800 r^3}
        +\frac{4969 p^6 (p.n)^2}{64 r^2}
\nonumber\\ &&      
  +\frac{275 p^8 (p.n)^2}{256 r}
        -\frac{4989943687 (p.n)^4}{352800 r^4}
        +\frac{2674877 p^2 (p.n)^4}{7840 r^3}
        +\frac{925 p^4 (p.n)^4}{24 r^2}
        +\frac{15 p^6 (p.n)^4}{128 r}
\nonumber\\ &&    
    -\frac{25649 (p.n)^6}{3360 r^3}
        -\frac{8331 p^2 (p.n)^6}{160 r^2}
        +\frac{751 (p.n)^8}{28 r^2}
\Biggr]
+\nu ^3 \Biggl[
        \frac{1617 p^{12}}{1024}
        -\frac{238966727 p^2}{151200 r^5}
\nonumber\\ &&    
    +\frac{127702733 p^4}{84672 r^4}
        +\frac{108551131 p^6}{4233600 r^3}
        +\frac{16283 p^8}{256 r^2}
        +\frac{3995 p^{10}}{256 r}
        +\pi ^2 \Biggl(
                -\frac{2339 p^2}{192 r^5}
                +\frac{98447 p^4}{3072 r^4}
\nonumber\\ &&              
  -\frac{20259 p^6}{1024 r^3}
                -\frac{16111 (p.n)^2}{192 r^5}
                +\frac{131231 p^2 (p.n)^2}{1536 r^4}
                +\frac{106947 p^4 (p.n)^2}{1024 r^3}
                -\frac{361499 (p.n)^4}{1024 r^4}
\nonumber\\ &&              
  -\frac{30075 p^2 (p.n)^4}{1024 r^3}
                -\frac{65625 (p.n)^6}{1024 r^3}
        \Biggr)
        +\frac{758233181 (p.n)^2}{151200 r^5}
        -\frac{10374288811 p^2 (p.n)^2}{705600 r^4}
\nonumber\\ &&      
  -\frac{2207947669 p^4 (p.n)^2}{1411200 r^3}
        +
        \frac{177 p^6 (p.n)^2}{256 r^2}
        -\frac{221 p^8 (p.n)^2}{64 r}
        +\frac{12810612439 (p.n)^4}{705600 r^4}
\nonumber\\ &&      
  +\frac{355111837 p^2 (p.n)^4}{94080 r^3}
        -\frac{125225 p^4 (p.n)^4}{768 r^2}
        -\frac{3 p^6 (p.n)^4}{128 r}
        -\frac{13905527 (p.n)^6}{4480 r^3}
\nonumber\\ &&      
  +\frac{136977 p^2 (p.n)^6}{1280 r^2}
        -\frac{15 p^4 (p.n)^6}{128 r}
        -\frac{289839 (p.n)^8}{4480 r^2}
        -\frac{35 p^2 (p.n)^8}{256 r}
\Biggr]
+\nu ^4 \Biggl[
        -\frac{1155 p^{12}}{1024}
\nonumber\\ &&       
 -\frac{593 p^6}{32 r^3}
        +\frac{6649 p^8}{256 r^2}
        -\frac{1615 p^{10}}{256 r}
        +\frac{549 p^4 (p.n)^2}{32 r^3}
        -\frac{62143 p^6 (p.n)^2}{256 r^2}
        +\frac{867 p^8 (p.n)^2}{256 r}
\nonumber\\ &&      
  -\frac{5749 p^2 (p.n)^4}{96 r^3}
        +\frac{652381 p^4 (p.n)^4}{768 r^2}
        -\frac{3 p^6 (p.n)^4}{64 r}
        -\frac{17623 (p.n)^6}{240 r^3}
        -\frac{1178329 p^2 (p.n)^6}{1280 r^2}
\nonumber\\ &&       
 -\frac{45 p^4 (p.n)^6}{128 r}
        +\frac{1443091 (p.n)^8}{4480 r^2}
        +\frac{105 p^2 (p.n)^8}{128 r}
\Biggr]
+\nu ^5 \Biggl[
        \frac{231 p^{12}}{1024}
        -\frac{63 p^{10}}{256 r}
        -\frac{35 p^8 (p.n)^2}{256 r}
\nonumber\\ &&
        -\frac{15 p^6 (p.n)^4}{128 r}
        -\frac{15 p^4 (p.n)^6}{128 r}
        -\frac{35 p^2 (p.n)^8}{256 r}
        -\frac{63 (p.n)^{10}}{256 r}
\Biggr]
+\frac{17 (p.n)^2}{4 r^5}
+\frac{29 p^2 (p.n)^2}{8 r^4}
\nonumber\\ &&
+\frac{21 p^4 (p.n)^2}{16 r^3}
+\frac{5 p^6 (p.n)^2}{32 r^2}
-\frac{(p.n)^4}{8 r^4}~.
\end{eqnarray}
%----------------------------------------------------------------------------------------------------------------
By this we have shown in explicit form the cancellation of the singularities originally occurring in harmonic 
coordinates, for reasons of {\it regularization} only. In the case of the binary point--mass problem up to 5PN
order no singularities survive requiring another method to be removed. At 4PN this has also been shown in 
Ref.~\cite{Blumlein:2020pog}, see also \cite{Porto:2017dgs}. Furthermore, logarithmic terms do now only occur
at $O(\nu)$ and $O(\nu^2)$.
%----------------------------------------------------------------------------------------------------------------
\section{Comparison to the literature}
\label{sec:5}
%----------------------------------------------------------------------------------------------------------------

\vspace*{-1mm}
In the following we perform a series of comparisons with results in the literature.
%----------------------------------------------------------------------------------------------------------------
\subsection{Canonical transformation to EOB}
\label{sec:51}
%----------------------------------------------------------------------------------------------------------------

\vspace*{1mm}
\noindent
Let us first compare to the EOB results of Ref.~\cite{Bini:2020wpo},~Eq.~(11.8), for the contributions at 
$O(\nu^0)$ and $O(\nu^3)$ and higher given in EOB coordinates in complete form.\footnote{For definiteness we use 
the minimal choice (8.24) of the flexibility parameters.} These terms do not receive contributions due to tail 
terms and one can therefore just refer to the pole--free Hamiltonian of Section~\ref{sec:4} to construct the canonical 
transformation.

It is given by
%----------------------------------------------------------------------------------------------------------------
\begin{eqnarray}
G(p^2,p.n,r) &=& p.n \Biggr\{t_1 \Biggl\{
        \nu  \Biggl(
                -\frac{1}{2}
                +\frac{1}{2} p^2 r
        \Biggr)
        - 1
\Biggr\}
+ t_2 \Biggl\{
        \nu  \Biggl(
                -\frac{5}{4} p^2 
                +\frac{5}{4 r}
                -\frac{1}{8} p^4 r
\nonumber\\ &&  
              -\frac{1}{2} (p.n)^2
        \Biggr)
        +\nu ^2 \Biggl(
                \frac{1}{4} p^2
                -\frac{1}{4 r}
                -\frac{1}{8} (p.n)^2
        \Biggr)
\Biggr\}
+ t_3 \Biggl\{
        \nu  \Biggl(
                \frac{\frac{29}{6} p^2 
                +\frac{4}{3} (p.n)^2
                }{r}
\nonumber\\ &&
                +\frac{7}{16} p^4 
                +\frac{
                        1795-63 \pi ^2}{72 r^2}
                +\frac{1}{16} p^6 r 
                +\frac{1}{4} p^2 (p.n)^2
        \Biggr)
\nonumber\\ &&
        +\nu ^2 \Biggl(
                \frac{-\frac{19}{24} p^2 
                -\frac{83}{24} (p.n)^2
                }{r}
                -\frac{37}{96} p^4 
                -\frac{3}{16 r^2}
                -\frac{1}{16} p^6 r 
                +\frac{7}{24} p^2 (p.n)^2
\nonumber\\ && 
                -\frac{1}{4} (p.n)^4 \Biggr)
+        \nu ^3 \Biggl(
                \frac{\frac{5}{24} p^2 
                -\frac{1}{8} (p.n)^2
                }{r}
                -\frac{1}{48} p^4 
                -\frac{3}{16 r^2}
                +\frac{1}{96} p^2 (p.n)^2
        \Biggr)
\Biggr\}
\nonumber\\ &&
+ t_4 \Biggl\{
        \nu ^3 \Biggl(
                \frac{1}{r} \Biggl[
                        -\frac{143}{16} p^4 
                        +\frac{493}{32} p^2 (p.n)^2
                        +\frac{31}{80} (p.n)^4
                \Biggr]
                +\frac{-\frac{15}{64} p^2 
                -\frac{15}{64} (p.n)^2
                }{r^2}
\nonumber\\ &&
                +\frac{229}{384} p^6 
                +\frac{7}{96 r^3}
                -\frac{1}{48} p^8 r 
                +\frac{61}{384} p^4 (p.n)^2
                -\frac{1}{16} p^2 (p.n)^4
        \Biggr]
\nonumber\\ &&
        +\nu^4 \Biggl[
                \frac{1}{r} \Biggl(
                        -\frac{1}{24} p^4 
                        +\frac{1}{48} p^2 (p.n)^2
                        +\frac{1}{48} (p.n)^4
                \Biggr)
                +\frac{\frac{5}{24} p^2 
                -\frac{1}{8} (p.n)^2
                }{r^2}
                -\frac{1}{6 r^3}
\nonumber\\ &&
                +
                \frac{1}{96} p^4 (p.n)^2
                -\frac{7}{128} p^2 (p.n)^4
                +\frac{5}{128} (p.n)^6
        \Biggr]
\Biggr\}
\nonumber\\ &&
+t_5 \Biggl\{
        \nu ^3 \Biggl[
                \frac{1}{r^3} \Biggl(
                        -\frac{p^2 \big(
                                -202645909+1786050 \pi ^2\big)}{100800}
\nonumber\\ && 
                        +\frac{\big(
                                -84723358+562625 \pi ^2\big) (p.n)^2}{33600}
                \Biggr)
                +\frac{1}{r^2} \Biggl(
                        -\frac{p^4 \big(
                                502480088+75854205 \pi ^2\big)}{6773760}
\nonumber\\ &&                       
 +\frac{p^2 \big(
                                -2251644296+120889125 \pi ^2\big) (p.n)^2}{5644800}
                        +\frac{\big(
                                17505304+16879275 \pi ^2\big) (p.n)^4}{2257920}
                \Biggr)
\nonumber\\ &&
                +\frac{1}{r} \Biggl(
                        \frac{7447}{128} p^6
                        +\frac{235}{12} p^4 (p.n)^2
                        -\frac{21323}{384} p^2 (p.n)^4
                        +\frac{4937}{420} (p.n)^6
                \Biggr)
                +\frac{595 p^8}{1536}
\nonumber\\ &&                
+\frac{
                        -186973+12400 \pi ^2}{640 r^4}
                -\frac{5}{48} p^{10} r
                -\frac{347}{384} p^6 (p.n)^2
                +\frac{109}{256} p^4 (p.n)^4
\nonumber\\ &&              
  -\frac{155}{256} p^2 (p.n)^6
        \Biggr]
        +\nu ^4 \Biggl[
                \frac{1}{r^2} \Biggl(
                        -\frac{7121}{640} p^4 
                        +\frac{105479 p^2 (p.n)^2}{5760}
                        -\frac{7073 (p.n)^4}{1440}
                \Biggr]
\nonumber\\ &&
                +\frac{1}{r} \Biggl(
                        \frac{14675}{384} p^6 
                        -\frac{100025 p^4 (p.n)^2}{1152}
                        +\frac{484729 p^2 (p.n)^4}{5760}
      -\frac{63677 (p.n)^6}{2688}
                \Biggr)
\nonumber\\ &&                  
                +\frac{-\frac{7}{40} p^2 
                +\frac{3829 (p.n)^2}{2880}
                }{r^3}
                -\frac{1927}{768} p^8 
                +\frac{13}{96 r^4}
                +\frac{145}{768} p^6 (p.n)^2
                +\frac{29}{128} p^4 (p.n)^4
\nonumber\\ &&              
  -\frac{203}{768} p^2 (p.n)^6
                +\frac{7}{48} (p.n)^8
        \Biggr]
        +\nu ^5 \Biggl[
                \frac{1}{r^2} \Biggl(
                        -\frac{77 p^4}{1280}
                        +\frac{23}{720} p^2 (p.n)^2
                        +\frac{697 (p.n)^4}{11520}
                \Biggr)
\nonumber\\ &&
                +\frac{1}{r} \Biggl(
                        -\frac{1}{384} p^6 
                        +\frac{13}{576} p^4 (p.n)^2
                        -\frac{199 p^2 (p.n)^4}{1440}
                        +\frac{13}{128} (p.n)^6
                \Biggr)
  + \frac{1}{r^3} \Biggl(\frac{107}{480} p^2 
\nonumber\\ &&              
                -\frac{377 (p.n)^2}{2880}\Biggr)
                +\frac{1}{768} p^8 
                -\frac{31}{192 r^4}
                +\frac{p^6 (p.n)^2}{4608}
                +\frac{37 p^4 (p.n)^4}{7680}
   -\frac{7 p^2 (p.n)^6}{1536}
        \Biggr]
\Biggr\}
\Biggr\}.
\end{eqnarray}
%----------------------------------------------------------------------------------------------------------------
In this way we confirm all the contributions of $O(\nu^0)$ and $O(\nu^3)$ or higher given in \cite{Bini:2020wpo} 
by an explicit Feynman diagram calculation ab initio.

%----------------------------------------------------------------------------------------------------------------
\subsection{The non-local terms}
\label{sec:52}
%----------------------------------------------------------------------------------------------------------------

\vspace*{1mm}
\noindent
Next we turn to the non--local terms defined in \cite{Bini:2020wpo}, cf.~Eq.~(\ref{eq:lonl}). 
We perform the eccentricity expansion of the non--local contributions for
$\langle \delta H_{\rm 4+5PN}^{\rm nl} \rangle$ with
%----------------------------------------------------------------------------------------------------------------
\begin{eqnarray}
\frac{\langle \delta H_{\rm 4+5PN}^{\rm nl} \rangle}{M c^2} &=& \frac{n}{2\pi M c^2} \int_0^{\tfrac{2\pi}{n}} dt~\delta H(t) 
\equiv 
F_{\rm 4+5PN}(a_r,e_t)
\\ 
&=& 
\frac{\nu^2}{a_r^5} \left[{\cal A}_{\rm 4PN} + {\cal B}_{\rm 4PN} \ln(a_r) \right] + 
\frac{\nu^2}{a_r^6} \left[{\cal A}_{\rm 5PN} + {\cal B}_{\rm 5PN} \ln(a_r) \right]
\end{eqnarray}
%----------------------------------------------------------------------------------------------------------------
starting from harmonic coordinates. Here $a_r$ is the semimajor axis of the orbit, which we rescaled by 
$a_r = a_{r, \rm phys} c^2/(G_N M)$. It 
appears in the parameterization of the radial 
coordinate distance $r$ in the form $r = a_r [1 - e_r \cos(u)]$, where $e_r$ denotes the ``radial eccentricity'' of the 
orbit and $u$ the ``eccentric anomaly''. The Kepler equation reads $n \cdot t =  1 - e_t \sin(u)$, with $n = 2\pi/P$. 
Here $P$ is the orbital period and $t$ the coordinate time defines the eccentricity $e_t$ and one uses standard relations
otherwise, cf.~\cite{BC}.

In the limit of vanishing eccentricity $e_t$ we obtain the following contribution for $\langle \delta H_{\rm 5PN}^{\rm nl} 
\rangle$, Eq.~(2.12), \cite{Bini:2020wpo}, 
%----------------------------------------------------------------------------------------------------------------
\begin{eqnarray}
\frac{\langle \delta {H}_{\rm 4+5PN}^{\rm nl} \rangle}{M c^2} 
&=& \frac{\nu^2}{a_r^5} \Biggl\{
        -\frac{32}{5} (\ln(a_r) - 2\gamma_E)
        +\frac{128}{5} \ln(2) \Biggr\}
%---
+\frac{\nu^2}{a_r^6}\Biggl\{
\Biggl(\frac{5854}{105} + \frac{56}{5} \nu \Biggr)(\ln(a_r) - 2 \gamma_E) 
\nonumber\\ &&
- \Biggl(\frac{25276}{105}
- \frac{912}{35} \nu\Biggr) \ln(2) 
+ \Biggl(\frac{243}{14} 
- \frac{486}{7} 
\nu\Biggr) \ln(3) + \frac{32}{5} \nu - \frac{96}{5} \Biggr\}, 
\end{eqnarray}
%----------------------------------------------------------------------------------------------------------------
which agrees with \cite{Bini:2020wpo}. The terms up to $O(e_t^{20})$ are given in Appendix~\ref{sec:B}. They agree
with the expansion coefficients of Table~I of \cite{Bini:2020wpo} up to $O(e_t^{10})$ (in the harmonic gauge).\footnote{
In Ref.~\cite{Bini:2020wpo} also the corresponding expressions in EOB coordinates are discussed.} 
The non--local contribution
to the energy for circular orbits is then obtained by\footnote{Note a difference to Eq.~(8.27), \cite{Bini:2020hmy}
in the $\ln(2) \nu^2$ term at 5PN.}
%----------------------------------------------------------------------------------------------------------------
\begin{eqnarray}
\frac{E^{\rm circ}_{\rm nl}}{\mu c^2} &=& \nu \Biggl\{
\Biggl[
        -\frac{64}{5} (\ln(j) -\gamma_E) 
        +\frac{128}{5} \ln(2) \Biggr] \frac{\eta^8}{j^{10}} 
+
\Biggl[
        \frac{32}{5}
        +\frac{28484}{105} \ln(2) 
        +\nu  \Biggl(
                \frac{32}{5}+\frac{112}{5} (\ln(j) 
\nonumber\\ &&
- \gamma_E) 
                +\frac{912}{35} \ln(2) 
- \frac{486}{7} \ln(3) \Biggr)
        +\frac{243}{14} \ln(3)
        -\frac{15172}{105} (\ln(j) - \gamma_E)
\Biggr] \frac{\eta^{10}}{j^{12}}
\Biggr\},
\end{eqnarray}
%----------------------------------------------------------------------------------------------------------------
with
%----------------------------------------------------------------------------------------------------------------
\begin{eqnarray}
\label{eq:ar1}
a_r = j^2 - 4 \eta^2 + O\left(\eta^4 \right),
\end{eqnarray}
%----------------------------------------------------------------------------------------------------------------
with $j = J_{\rm phys} c/(G_N M)$. Here we have also introduced the dimensionless quantity $\eta^2$, accounting for $1/c^2$.

The contributions up to $O(e_t^2)$ are needed below to derive periastron advance for 
circular motion.  One may now further express the variables $a_r$ and $e_t$ in terms of the normalized Delaunay 
variables \cite{DELAUNAY} $i_r, i_\phi$ and $i_{r \phi}$,  cf.~\cite{Bini:2020hmy}, 
Eq.~(A11), with $i_{r\phi} = i_r + i_\phi$, $i_\phi = j$. By this one obtains
%----------------------------------------------------------------------------------------------------------------
\begin{eqnarray}
\hat{F}(i_r,j) = F(a_r,e_t)~.
\end{eqnarray}
%----------------------------------------------------------------------------------------------------------------
The variables $\hat{H}, i_r$  and $i_\phi$ are related by Euler's chain rule
%----------------------------------------------------------------------------------------------------------------
\begin{eqnarray}
\label{eq:CHAIN}
\left.\frac{\partial \hat{H}}{\partial i_\phi}\right|_{i_r}
\left.\frac{\partial i_\phi}{\partial i_r}\right|_{\hat{H}}
\left.\frac{\partial i_r}{\partial \hat{H}}\right|_{i_\phi} = - 1
\end{eqnarray}
%----------------------------------------------------------------------------------------------------------------
since $\hat{H}$ depends only on $i_r$ and $i_\phi$ and therefore a function $f$ exists with $f(\hat{H}, i_r, i_\phi) = 0$.
By applying the chain rule one obtains the periastron advance, $K$, defined in (\ref{eq:PER1a})\footnote{Note that also a 
related quantity, $k = K - 1$, is sometimes denoted by periastron advance.} 
One obtains
%----------------------------------------------------------------------------------------------------------------
\begin{eqnarray}
\label{eq:PER1a}
K        &=& \frac{1}{\Omega_R} \left.\frac{\partial \hat{H}(i_r,j)}{\partial i_\phi}\right|_{i_r}
\\
\label{eq:OM1}
\Omega_R &=& \left.\frac{\partial \hat{H}(i_r,j)}{\partial i_r}\right|_{j}. 
\end{eqnarray}
%----------------------------------------------------------------------------------------------------------------
Here $\hat{H}$ denotes the complete Hamiltonian. One may express
%----------------------------------------------------------------------------------------------------------------
\begin{eqnarray}
\label{eq:PER1}
K        &=&  K_{\rm loc} + K_{\rm nl},
\\
\hat{H}  &=&  \hat{H}_{\rm loc} + \hat{H}_{\rm nl},
\end{eqnarray}
%----------------------------------------------------------------------------------------------------------------
where $K_{\rm nl}$ starts at 4PN and $\Omega_R$ receives non--local (nl) contributions from 4PN on, cf.~(\ref{eq:OM1}).
The contributions to $K_{\rm loc}$ are calculated in Section~\ref{sec:53}.
For $K^{\rm nl}_{\rm 4+5PN}$ the 4PN non--local contributions to $\Omega_R$ are necessary beyond the 1PN (local) 
correction
%----------------------------------------------------------------------------------------------------------------
\begin{eqnarray}
\Omega_R^{\rm loc, 1PN}  &=& i_{r\phi}^{-3}\left[1 + \frac{(3+ \nu) i_\phi + 18 i_r}{2 i_{\phi} i_{r\phi}^2} \eta^2\right] + 
O\left(\eta^4\right),
\label{eq:OM3}
\end{eqnarray}
%----------------------------------------------------------------------------------------------------------------
\cite{Bini:2020hmy} with $\Omega_R = (G_N M/c^3) \Omega_{R, \rm phys}$. We first calculate $\Omega_R^{\rm nl, 4PN}$, setting 
$i_\phi = j$, for circular orbits
%----------------------------------------------------------------------------------------------------------------
\begin{eqnarray}
\Omega_R^{\rm nl, 4PN}  &=& \left.\frac{\partial \hat{H}^{\rm nl, 4PN}(i_r,j)}{\partial i_r}\right|_{j;~i_r = 0}
\nonumber\\ &=& -\frac{64}{10} \nu \frac{\eta^8}{j^{11}} \Bigg[13 + \frac{37}{6} ( \ln(j) -  \gamma_E)  + \frac{203}{6} \ln(2) - 
\frac{729}{16} \ln(3)\Biggr]. 
\label{eq:OM4}
\end{eqnarray}
%----------------------------------------------------------------------------------------------------------------
The Newtonian term of $\Omega_R$ for circular orbits is $1/j^3$. Since we are only considering the 4 and 5PN
contributions, the post--Newtonian expansion of $1/\Omega_R$ can be done separately for (\ref{eq:OM3}) and 
(\ref{eq:OM4}), keeping the Newtonian contribution. The second term hits the $O(\eta^2)$ term of $K_{\rm loc}$
(\ref{eq:KL1}). In this way Eq.~(8.21) in \cite{Bini:2020hmy} needs a slight extension.

We obtain
%----------------------------------------------------------------------------------------------------------------
\begin{eqnarray}
K^{\rm nl}_{\rm 4PN}(j) &=& -\frac{64}{10} \nu \frac{\eta^8}{j^8} \Biggl[ - 11 
- \frac{157}{6} (\ln (j) - \gamma_E) + \frac{37}{6} \ln(2) + \frac{729}{16} \ln(3) \Biggr]
\label{eq:NLP4}
\\
K^{\rm nl}_{\rm 5PN}(j) &=& -\frac{64}{10} \nu \frac{\eta^{10}}{j^{10}}
\Biggl[
-\frac{59723}{336} - \frac{9421}{28} [\ln(j) - \gamma_E] + \frac{7605}{28}  \ln(2)
+ \frac{112995}{224} \ln(3)
\nonumber\\ &&
+ \Biggl( 
  \frac{2227}{42} 
+ \frac{617}{6} [\ln(j) - \gamma_E] 
- \frac{7105}{6} \ln(2) 
+ \frac{54675}{112} \ln(3)
\Biggr) \nu 
\Biggr],
\label{eq:NLP5}
\end{eqnarray}
%----------------------------------------------------------------------------------------------------------------
which is calculated in a different way than the local contributions. The representations are, however, 
equivalent. Here one has 
%----------------------------------------------------------------------------------------------------------------
\begin{eqnarray}
\label{eq:ar2}
a_r       &=& i_{r \phi}^2 - 2 \frac{3 i_r +2 i_\phi}{i_\phi} \eta^2 + O\left(\eta^4\right),
\\
e_t^2     &=& \frac{i_r}{i_{r\phi}^2} \left[ i_r + 2 i_\phi + 2 \frac{i_r(\nu-1) + i_\phi (2 \nu-5)}{i_{r\phi}^2} 
\eta^2  
\right] + O\left(\eta^4\right),
\end{eqnarray}
%----------------------------------------------------------------------------------------------------------------
cf.~\cite{Bini:2020hmy}.
Eq.~(\ref{eq:ar2}) turns into (\ref{eq:ar1}) for circular orbits ($i_r \rightarrow 0$).
Eq.~(\ref{eq:NLP4}) agrees with Eq.~(5.7) in \cite{Bernard:2016wrg}, see also the expression of the related 
function $\rho(x)$ in \cite{Damour:2016abl} and Eqs.~(\ref{eq:NLP4},\ref{eq:NLP5}) agree with Eq.~(8.29) of 
\cite{Bini:2020hmy}.
%----------------------------------------------------------------------------------------------------------------
\subsection{Periastron advance: local terms}
\label{sec:53}
%----------------------------------------------------------------------------------------------------------------

\vspace*{1mm}
\noindent 
The local contribution to periastron advance is obtained by
%----------------------------------------------------------------------------------------------------------------
\begin{eqnarray}
\label{eq:INTperi}
K_{\rm loc} = - \frac{1}{\pi}\frac{\partial}{\partial j} \int_{r_{\rm min}}^{r_{\rm max}} dr 
\sqrt{R(r,\hat{E},j)}.
\end{eqnarray}
%----------------------------------------------------------------------------------------------------------------
Here $\hat{E}$ results from (\ref{eq:HatH}) by $H \rightarrow E$ and 
%----------------------------------------------------------------------------------------------------------------
\begin{eqnarray}
\left. R(r,\hat{E},j)\right|_{\rm 5PN} 
= A +  \frac{2B}{r} + \frac{C}{r^2} + \eta^2 \frac{D_1}{r^3}
+ \sum_{k = 1}^4 \eta^{2(k+1)} \left[ \frac{D_{2k}}{r^{2k+2}} + \frac{D_{2k+1}}{r^{2k+3}} \right].
\label{eq:RES}
\end{eqnarray}
%----------------------------------------------------------------------------------------------------------------
It is convenient to refer to the local terms rather than to a separation of potential and tail terms. The former ones
have no logarithmic terms and the corresponding integrals are therefore somewhat simpler. The logarithmic terms 
have already been dealt with in Section~\ref{sec:52}. 

The integrand of (\ref{eq:INTperi}) has the form
%----------------------------------------------------------------------------------------------------------------
\begin{eqnarray}
\sqrt{R(r,\hat{E},j)} = 
\frac{1}{r}\sqrt{A r^2 + 2 B r  + C} + \eta^2 \frac{D_1}{2 r^2 \sqrt{A r^2 + 2 B r  + C}} + 
O\left( \eta^4 \right).
\end{eqnarray} 
%----------------------------------------------------------------------------------------------------------------
 The relation  
%----------------------------------------------------------------------------------------------------------------
\begin{eqnarray}
\hat{E} = \hat{H}(p^2,(p.n)^2,r) 
\label{eq:pn2} 
\end{eqnarray}
%----------------------------------------------------------------------------------------------------------------
is solved iteratively for $R(r,\hat{E},j) = (p.n)^2$ by applying 
%----------------------------------------------------------------------------------------------------------------
\begin{eqnarray}
p^2 = (p.n)^2 + \frac{j^2}{r^2},
\end{eqnarray}
%----------------------------------------------------------------------------------------------------------------
through which the functions $A, B, C$ and $D_k$ become polynomials in $\hat{E}$ and $j$. The integral (\ref{eq:INTperi}) is
usually solved by a mapping to a contour integral \cite{SOMMBORN} applying the residue theorem, expanding
in $\eta^2$ up to 5PN. Except of the integral for the Newtonian term, involving only $A,B$ and $C$, all other integrals 
have only one residue at $r=0$, see Appendix~\ref{sec:C}.

We calculate the local contribution to periastron advance starting from harmonic coordinates and compare to 
Eq.~(F5) of \cite{Bini:2020hmy} resulting from the local EOB Hamiltonian Eq.~(11.8) \cite{Bini:2020wpo}.
This is necessary to fix the notion of the parameters $\bar{d}_5$ and $a_6$ in $K(E,j)_{\rm loc,f}$ to 5PN. 
We rather use  $K(E,j)_{\rm loc,f}$ than $K_{\rm loc,5PN}^{\rm circ}$ to test three relations between the 
parameters $\bar{d}_5$ and $a_6$, which is advantageous. 

To 4PN one obtains
%----------------------------------------------------------------------------------------------------------------
\begin{eqnarray}
K(\hat{E},j)_{\rm loc,f}^{\rm \leq 4PN} &=& 1 + \frac{3}{j^2} \eta^2 
+ \Biggl[
  \Biggl( \frac{15}{2} - 3 \nu\Biggr) \frac{\hat{E}}{j^2} + \Biggl( \frac{105}{4} - \frac{15 \nu}{2} \Biggr) 
  \frac{1}{j^4} 
  \Biggr] \eta^4
+ \Biggl[
  \Biggl(\frac{15}{4}(1-\nu) + 3 \nu^2\Biggr) \frac{\hat{E}^2}{j^2}
\nonumber\\ &&
   + \Biggl(
\frac{315}{2} 
+ \Biggl(\frac{123 \pi^2}{64} - 218\biggr) \nu 
+\frac{45 }{2} \nu^2 
\Biggr) \frac{\hat{E}}{j^4}
+ \Biggl(
	 \frac{1155}{4} 
+ \Biggl(\frac{615 \pi^2}{128} 
- \frac{625}{2} \Biggr) \nu 
\nonumber\\ &&
+\frac{105}{8} \nu^2 \Biggr)  
\frac{1}{j^6} \Biggr] \eta^6      
+\Biggl[
\Biggl(\frac{15}{4}-3 \nu\Biggr) \nu^2 \frac{ \hat{E}^3}{j^2}
+\Biggl(
\frac{4725}{16}
+\Biggl(\frac{35569 \pi ^2}{2048}-\frac{20323}{24} \nu 
\nonumber\\ &&
+\Biggl(\frac{4045}{8}-\frac{615 \pi ^2}{128}\Biggr) \nu^2
-45 \nu^3
\Biggr)
\frac{ \hat{E}^2}{j^4}  +\Biggl(-\frac{525 \nu ^3}{8}+\Biggl(\frac{35065}{16}
-\frac{615 \pi^2}{16}\Biggr) \nu^2
\nonumber\\ &&
+\Biggl(\frac{257195 \pi^2}{2048}
-\frac{293413}{48}\Biggr) \nu +\frac{45045}{16}\Biggr)\frac{ \hat{E}}{j^6}
+\Biggl(
\frac{225225}{64}
+\Biggl(\frac{2975735 \pi ^2}{24576}
\nonumber\\ &&
-\frac{1736399}{288}\Biggr) \nu 
+\Biggl(\frac{132475}{96}-\frac{7175 \pi ^2}{256}\Biggr) \nu^2
-\frac{315}{16} \nu^3
\Biggr)
\frac{1}{j^8} \Biggr] \eta^8.
\label{eq:KL1}
\end{eqnarray}
%----------------------------------------------------------------------------------------------------------------
We also rederived the periastron advance starting with the ADM Hamiltonian \cite{Damour:2014jta} and confirm the 
result given in \cite{Bernard:2016wrg,Bini:2020hmy}.
For the 5PN terms we obtain the partial result 
%----------------------------------------------------------------------------------------------------------------
\begin{eqnarray}
K(\hat{E},j)_{\rm loc,f}^{\rm 5PN} &\propto& 
%----
\Biggl\{
\Biggl[
\frac{15 \nu^2}{16}
 - \frac{15}{4} \nu^3
 + 3 \nu^4
\Biggr]\frac{ \hat{E}^4}{j^2}  
%--
+\Biggl[
\frac{3465}{16}
+\frac{15829 \pi ^2}{256} \nu
-\frac{35569 \pi ^2}{1024} \nu^2
+\Biggl(\frac{1107 \pi ^2}{128}
\nonumber\\ &&
-\frac{7113}{8}\Biggr) \nu^3
+75\nu ^4
\Biggr]
\frac{ \hat{E}^3}{j^4}
%---
+\Biggl[
 \frac{315315}{32}
+\frac{4899565 \pi ^2}{4096} \nu
-\frac{3289285 \pi^2}{1024}  \nu^2
+\Biggl(\frac{35055 \pi ^2}{256}
\nonumber\\ &&
-\frac{240585}{32}\Biggr) \nu^3 
+\frac{1575}{8} \nu^4
\Biggr)
\frac{ \hat{E}^2}{j^6} 
%---
+\Biggl[ \frac{765765}{16} 
+ \frac{16173395 \pi ^2}{8192} \nu 
- \frac{77646205 \pi^2}{8192} \nu^2
\nonumber\\ &&
+ \Biggl(
\frac{121975 \pi ^2}{512}
-\frac{271705}{24}\Biggr) \nu^3 
+\frac{2205}{16} \nu^4
\Biggr]
\frac{ \hat{E}}{j^8} 
+\Biggl[
 \frac{2909907}{64}
+ \frac{1096263 \pi^2}{1024} \nu
\nonumber\\ &&
-\frac{87068961 \pi^2}{16384} \nu^2
+\Biggl(\frac{90405 \pi^2}{1024}-\frac{127995}{32}\Biggr) \nu^3
+\frac{3465}{128} \nu^4
\Biggr]
\frac{1}{j^{10}}
\Biggr\} \eta^{10},
\label{eq:PERI2}
\end{eqnarray}
%----------------------------------------------------------------------------------------------------------------
leaving out the few rational terms at $O(\nu)$ and $O(\nu^2)$ still to be calculated in complete form. The terms given in
(\ref{eq:PERI2}) agree with those of Ref.~\cite{Bini:2020hmy} considering our results in (\ref{eq:d5}) and (\ref{eq:a6}).

Eq.~(\ref{eq:PERI2}) allows to derive the $\pi^2$ contributions to $\bar{d}_5$ and $a_6$ to which we turn now.
%----------------------------------------------------------------------------------------------------------------
\subsection{The \boldmath $\pi^2$ contributions}
\label{sec:54}
%----------------------------------------------------------------------------------------------------------------

\vspace*{1mm}
\noindent
The $\pi^2$--contributions stem from the potential terms. They also contribute to the yet open constants $\bar{d}_5$ 
and $a_6$ in \cite{Bini:2020wpo} at $O(\nu^2)$. They can be extracted from the corresponding contribution to the binding 
energy for circular motion (for $a_6$) and the circular periastron 
advance\cite{Damour:1988mr,Damour:1999cr,Bernard:2016wrg}, 
respectively. One obtains
%----------------------------------------------------------------------------------------------------------------
\begin{eqnarray}
\label{eq:d5}
\bar{d}_5  &=& r_{\bar{d}_5} + \frac{306545}{512} \pi^2
\\
\label{eq:a6}
{a}_6  &=& r_{{a}_6}     + \frac{25911}{256} \pi^2
\end{eqnarray}
%----------------------------------------------------------------------------------------------------------------
from Eqs.~(\ref{eq:PERI2},\ref{Ecirc},\ref{Kcirc}). We also agree with the $\pi^2$ contributions of 
$O(\nu)$. Let us finally summarize the terms for the binding energy $E^{\rm circ}(j)$ and the periastron advance 
$K^{\rm circ}(j)$ at circular orbits obtained in the present calculation,
%----------------------------------------------------------------------------------------------------------------
\begin{eqnarray}
\label{Ecirc}
\frac{E^{\rm circ}(j)}{\mu c^2} 
&=& -\frac{1}{2j^2} 
%--
+\left(-\frac{\nu }{8}-\frac{9}{8}\right)\frac{1}{j^4} \eta^2
%--
+\left(-\frac{\nu ^2}{16}+\frac{7 \nu }{16}-\frac{81}{16}\right)\frac{1}{j^6} \eta^4
%--
+\Biggl[-\frac{5 \nu ^3}{128}+\frac{5 \nu ^2}{64}+\left(\frac{8833}{384}
\right. \nonumber\\ && \left.
-\frac{41 \pi ^2}{64}\right) \nu 
-\frac{3861}{128}
\Biggr] \frac{1}{j^8} \eta^6
%--
+
\Biggl[-\frac{7 \nu ^4}{256}+\frac{3 \nu ^3}{128}+\left(\frac{41 \pi
   ^2}{128}-\frac{8875}{768}\right) \nu
   ^2+\left(\frac{989911}{3840} \right.
\nonumber\\ && \left.
-\frac{6581 \pi ^2}{1024}\right) \nu
   -\frac{53703}{256}
\Biggr] \frac{1}{j^{10}} \eta^8
+
\Biggl[\left( 
 r^{\rm E}_{\nu^2}
+ \frac{132979 \pi^2}{2048} 
\right) \nu ^2-\frac{21 \nu^5}{1024}
+\frac{5 \nu ^4}{1024}
\nonumber\\ &&
+\left(\frac{41 \pi^2}{512}
-\frac{3769}{3072}\right) \nu^3
+\left(r^{\rm E}_{\nu}-\frac{31547 \pi ^2}{1536}\right) \nu 
-\frac{1648269}{1024} \Biggr] \frac{1}{j^{12}} \eta^{10} 
%\nonumber\\ &&
+ \frac{E^{\rm circ}_{\rm nl}}{\mu c^2} + 
O\left( \eta^{12} \right),
\nonumber\\
%----
\\
\label{Kcirc}
K^{\rm circ}(j)
&=&
1 + 3 \frac{1}{j^2} \eta^2 
%--
+\left(\frac{45}{2}-6\nu\right)\frac{1}{j^4} \eta^4
%--
+\left[\frac{405}{2}+\left(-202+\frac{123}{32}\pi^2\right)\nu+3\nu^2\right]\frac{1}{j^6} \eta^6
%--
\nonumber\\ &&
+
\left[\frac{15795}{8}+\left(\frac{185767}{3072}\pi^2-\frac{105991}{36}\right)\nu
+\left(-\frac{41}{4}\pi^2+\frac{2479}{6}\right)\nu^2\right]\frac{1}{j^8} \eta^8 
%--
+
\Biggl[\frac{161109}{8}
\nonumber\\ && 
+\left(r_{\nu}^{\rm K} +\frac{488373}{2048}\pi^2\right)\nu
+\left(r_{\nu^2}^{\rm K}  - \frac{1379075}{1024} \pi^2
\right)\nu^2
\nonumber\\ &&
+\left(-\frac{1627}{6}+\frac{205}{32}\pi^2\right)\nu^3\Biggr]\frac{1}{j^{10}} \eta^{10} 
+ K^{\rm nl}_{\rm 4+5PN}(j) + O\left(\eta^{12} \right),
\end{eqnarray}
%----------------------------------------------------------------------------------------------------------------
cf. also \cite{Damour:1999cr}, Eq.~(5.25), and \cite{Bernard:2016wrg}, Eq.~(5.9).
Note that there are correlations between the rational quantities $r_{\nu,\nu^2}^{\rm E,K}$ and $r_{\bar{d}_5,a_6}$. 

To obtain the corresponding relations in terms of the variable $x = (G_N M \Omega_\phi/c^3)^{2/3}$, with $\Omega_\phi$ 
the angular frequency, one may apply $j = j(x)$, Eq.~(8.31) of \cite{Bini:2020hmy}. In particular one has in the Schwarzschild
limit of (\ref{Ecirc}, \ref{Kcirc})
%----------------------------------------------------------------------------------------------------------------
\begin{eqnarray}
\label{Esch}
\frac{E^{\rm Schw.}(x)}{\mu c^2} &=& \frac{1-2x}{\sqrt{1-3x}} - 1,
\\
\label{Ksch}
K^{\rm Schw.}(x)            &=& \frac{1}{\sqrt{1-6x}},
\end{eqnarray}
%----------------------------------------------------------------------------------------------------------------
where (\ref{Esch}) has been given in\cite{Schafer:2018kuf}. The relation for $K^{\rm Schw.}(j)$ has been given in 
\cite{Damour:1988mr}, Eq.~(A8).

%----------------------------------------------------------------------------------------------------------------
\subsection{Comparison to the other contributions to the tail terms}
\label{sec:55}
%----------------------------------------------------------------------------------------------------------------

\vspace*{1mm}
\noindent
Let us mention that we have recalculated the contributions to the tail term given in \cite{Foffa:2019eeb}
in $D$ dimensions, but do not agree with all terms in the rational (local) contributions.
A further detailed comparison has to be performed, which will be given elsewhere. 

There are multi--pole moment contributions containing products of Levi--Civita tensors $\varepsilon_{ijk}$, which 
have to be dealt with in $D-1$ dimensions in the present approach, see also \cite{Blanchet:2005tk,Foffa:2019eeb}. 
Despite the fact that products of two
(or more) Levi--Civita symbols, here in Euclidean space, are turned into determinants of $D-1$ Kronecker symbols
\cite{IZ} it is known from almost all applications in Quantum Field Theory, that a so--called finite 
renormaliztion has to be performed to re-establish the Ward--identities.\footnote{In the present application to 
gravity one should not call this `finite renormalization', since no renormalization is performed. However,
the contribution to the observables will be different comparing the 3--dimensional properly regulated result
with that obtained in $D-1$ dimensions.}  The Larin method 
\cite{Larin:1993tq} is one consistent way (i.e. a non--degenerative way) to perform this analytic continuation to 
$D$ dimensions.\footnote{Other prescriptions were given in \cite{GA5}.}  
Also this aspect still needs further study. 

We finally mention that for the circular binding energy and periastron advance the yet differing results in 
the tail terms yield a numerical effect on the difference of $O(1\%)$ or less. This remaining difference 
still has to be settled analytically.
%----------------------------------------------------------------------------------------------------------------
\section{Conclusions}
\label{sec:6}
%----------------------------------------------------------------------------------------------------------------

\vspace*{1mm}
\noindent
We have presented the 5PN potential contributions to the Hamiltonian of binary motion in gravity 
starting form the harmonic gauge and a part of the 5PN tail term. The calculation has thoroughly been performed 
in $D$ dimensions, based on 188533 Feynman diagrams using effective field theory methods, as a calculation ab initio.
The singular and logarithmic contributions to the 5PN tail terms have been calculated.
We have shown the explicit cancellation of the singularities between both contributions, performing an additional 
canonical transformation to a pole--free Hamiltonian. We have shown in an explicit calculation how to match the 
potential and the tail terms, using dimensional regularization. Here the overlap--terms are canceling.

Comparisons to the literature have been performed. Firstly, we have shown that all terms of $O(\nu^0)$ and $O(\nu^3)$ 
and higher agree with the results presented in the literature. At $O(\nu^2)$ we determined the $\pi^2$ contributions 
to $\bar{d}_5$ and $a_6$. Furthermore, we also agree with the logarithmic tail and potential terms and the  $\pi^2$--terms 
at $O(\nu)$ and the effect of the non--local terms on $E_{\rm circ,5PN}$ and $K_{\rm circ,5PN}$. We still observe a few 
differences in the purely rational (local) contributions to the tail term comparing to the present literature, which 
have to be clarified to obtain the complete 5PN result.

\appendix
%----------------------------------------------------------------------------------------------------------------
\section{Joining the potential and the tail term and the method of expansion by regions}
\label{sec:A}
%----------------------------------------------------------------------------------------------------------------

\vspace*{1mm}
\noindent
We now prove that the Taylor expansions in the potential and ultrasoft region commute and that the occurring 
overlap integrals are indeed scaleless and vanish in $D$ dimensions. Here we use arguments given in 
\cite{Jantzen:2011nz}.

The general form of the contributing integrands $I$ is 
%----------------------------------------------------------------------------------------------------------------
\begin{equation}
  \label{eq:I}
  I = \frac{\exp\big(i k[x(t_1) - x(t_2)]\big)}{k^2\prod_{i=1}^{P_{\text{us}}} (k + p_i)^2 
\prod_{i=1}^{P_{\text{pot}}} (k + q_i)^2 } \mathcal{J}\big(\{q_i\}, \{p_i\}\big) 
\mathcal{P}\big(k, \{q_i\}, \{p_i\}, v_1, v_2\big)\,.
\end{equation}
%----------------------------------------------------------------------------------------------------------------
Apart from the loop momentum $k$, $I$ depends on ultrasoft momenta $p_i, 1 \leq i \leq P_{\text{us}}$ and 
potential momenta $q_i, 1 \leq i \leq P_{\text{pot}}$. The loop is associated with one of the two worldlines, 
whose four--position at the time $t$ is given by $x(t)$. $\mathcal{J}$ is a function with the same structure as 
$I$ itself, but independent of $k$. Finally, $\mathcal{P}$ denotes a polynomial in the momenta and worldline 
velocities $v_1, v_2$. Our aim is to show that
%----------------------------------------------------------------------------------------------------------------
\begin{equation}
  \label{eq:T_commute}
  T_{\text{pot}}^N T_{\text{us}}^N I = T_{\text{us}}^N T_{\text{pot}}^N I 
   =  \sum_{i=0}^{i_{\rm max}(N)} \frac{\mathcal{J}_i \mathcal{P}_i}{\vec{k}^{2i}}  \,,
\end{equation}
%----------------------------------------------------------------------------------------------------------------
according to the power counting in the respective region. The $\mathcal{P}_i$ are polynomials in the components 
of the four--vectors appearing in Eq.~(\ref{eq:I}) and the $\mathcal{J}_i$ are
independent of $k$. This structure implies
%----------------------------------------------------------------------------------------------------------------
\begin{equation}
  \label{eq:int_I}
  \int_{\mathds{R}^d} d^d\vec{k}\ T_{\text{pot}}^N T_{\text{us}}^N I = 0\,.
\end{equation}
%----------------------------------------------------------------------------------------------------------------

We first note that
%----------------------------------------------------------------------------------------------------------------
\begin{equation}
  \label{eq:T_hom}
  T_{\text{pot}}^N f\,g = (T_{\text{pot}}^N \,f) (T_{\text{pot}}^N \,g)\,,
\end{equation}
%----------------------------------------------------------------------------------------------------------------
and similar for $T_{\text{us}}^N$. This allows us to expand each factor in Eq.~(\ref{eq:I}) separately. $\mathcal{J}$ is 
independent of $k$, and $\mathcal{P}$ is a polynomial, so trivially
%----------------------------------------------------------------------------------------------------------------
\begin{align}
  \label{eq:TT_J}
  T_{\text{pot}}^N T_{\text{us}}^N \mathcal{J} ={}& T_{\text{us}}^N T_{\text{pot}}^N \mathcal{J} = \mathcal{J}\,,\\
  \label{eq:TT_P}
  T_{\text{pot}}^N T_{\text{us}}^N \mathcal{P} ={}& T_{\text{us}}^N T_{\text{pot}}^N \mathcal{P} = \mathcal{P}\,.
\end{align}
%----------------------------------------------------------------------------------------------------------------
For the propagators without additional momenta, we obtain
%----------------------------------------------------------------------------------------------------------------
\begin{equation}
  \label{eq:TT_k2}
  T_{\text{pot}}^N T_{\text{us}}^N \frac{1}{k^2} = T_{\text{pot}}^N \frac{1}{k^2} = \sum_{n=0}^{\frac{N}{2}} \frac{k_0^{2n}}
{(k^2)^{n+1}} = T_{\text{us}}^N \sum_{n=0}^{\frac{N}{2}} \frac{k_0^{2n}}{(k^2)^{n+1}} = T_{\text{us}}^N 
T_{\text{pot}}^N \frac{1}{k^2}\,,
\end{equation}
%----------------------------------------------------------------------------------------------------------------
and similar
%----------------------------------------------------------------------------------------------------------------
\begin{align}
  \label{eq:TT_k+p}
  T_{\text{pot}}^N T_{\text{us}}^N \frac{1}{(k+p_i)^2} ={}& T_{\text{pot}}^N \frac{1}{(k+p_i)^2} = \sum_{n=0}^N 
\frac{T_\text{pot}^N \big[(k_0+p_{i0})^2 - 2\vec{k}\vec{p_i} - \vec{p}_i^{\,2}\big]^n}{(k^2)^{n+1}}\,,\\
T_{\text{us}}^N T_{\text{pot}}^N \frac{1}{(k+p_i)^2}
  ={}& T_{\text{us}}^N \sum_{n=0}^N \frac{T_\text{pot}^N \big[(k_0+p_{i0})^2 - 2\vec{k}\vec{p_i} - \vec{p}_i^{\,2}\big]^n}
{(k^2)^{n+1}} \notag\\
  ={}& \sum_{n=0}^N \frac{T_\text{pot}^N \big[(k_0+p_{i0})^2 - 2\vec{k}\vec{p_i} - 
\vec{p}_i^{\,2}\big]^n}{(k^2)^{n+1}}\,.
\end{align}
%----------------------------------------------------------------------------------------------------------------

With respect to the remaining propagators, we first observe the absence
of poles in $v$, i.e.
%----------------------------------------------------------------------------------------------------------------
\begin{equation}
  \label{eq:T_minus}
  T_{\text{pot}}^Z \frac{1}{(k+q_i)^2}
= T_{\text{pot}}^Z T_{\text{us}}^N \frac{1}{(k+q_i)^2}
= T_{\text{us}}^Z \frac{1}{(k+q_i)^2}
= T_{\text{us}}^Z T_{\text{pot}}^N \frac{1}{(k+q_i)^2}
= 0 \qquad \text{for all }Z < 0\,,
\end{equation}
%----------------------------------------------------------------------------------------------------------------
and note the following algebraic properties of the Taylor expansion operators.
%----------------------------------------------------------------------------------------------------------------
\begin{align}
  T_{\text{pot}}^N P ={}&  P\, T_{\text{pot}}^{N-1}\,, & P \in{}&\{k_0, p_{i0}, q_{i0}, \vec{p}_i\}\\
T_{\text{pot}}^N P ={}&  P\, T_{\text{pot}}^{N}\,, & P \in{}&\{\vec{k}, \vec{q}_i \}\\
  T_{\text{us}}^N P ={}&  P\, T_{\text{us}}^{N-1}\,, & P \in{}&\{k_0, p_{i0}, q_{i0}, \vec{k}, \vec{p}_i\}\\
T_{\text{us}}^N P ={}&  P\, T_{\text{us}}^{N}\,, & P \in{}&\{\vec{q}_i \}.
\end{align}
%----------------------------------------------------------------------------------------------------------------
We now show that
%----------------------------------------------------------------------------------------------------------------
\begin{equation}
  \label{eq:TT_pot_commute}
  T_{\text{pot}}^N T_{\text{us}}^M \frac{1}{(k+q_i)^2} =  T_{\text{us}}^M T_{\text{pot}}^N \frac{1}{(k+q_i)^2}
\end{equation}
%----------------------------------------------------------------------------------------------------------------
by induction over $N+M$. The case $N+M=0$ is straightforward. For $N+M
\geq 1$ we observe
%----------------------------------------------------------------------------------------------------------------
\begin{align}
  \label{eq:TpotTus_pot}
  T_{\text{pot}}^N T_{\text{us}}^M \frac{1}{(k+q_i)^2} ={}& \frac{T_{\text{pot}}^N}{\vec{q}_i^{\,2}}\left( 1 + \big[(k_0 + q_0)^2 -\vec{k}^2\big] T_{\text{us}}^{M-2} \frac{1}{(k+q_i)^2} - 2\,\vec{k}\vec{q}_i\,T_{\text{us}}^{M-1} \frac{1}{(k+q_i)^2}\right)\,,\notag\\
  ={}& \frac{1}{\vec{q}_i^{\,2}}\left( 1 + (k_0 + q_0)^2 T_{\text{pot}}^{N-2} T_{\text{us}}^{M-2} \frac{1}{(k+q_i)^2}  \right.\notag\\
  &\left. -\vec{k}^2 T_{\text{pot}}^{N}T_{\text{us}}^{M-2} \frac{1}{(k+q_i)^2} - 2\,\vec{k}\vec{q}\,T_{\text{pot}}^{N}\,T_{\text{us}}^{M-1} \frac{1}{(k+q_i)^2}\right)\,,
\end{align}
%----------------------------------------------------------------------------------------------------------------
and similar
%----------------------------------------------------------------------------------------------------------------
\begin{align}
  \label{eq:TusTpot_pot}
  T_{\text{us}}^M T_{\text{pot}}^N \frac{1}{(k+q_i)^2} ={}& \frac{T_{\text{us}}^M}{\vec{q}_i^{\,2}}\left( 1 + (k_0 + q_0)^2 T_{\text{pot}}^{N-2} \frac{1}{(k+q_i)^2} - (\vec{k}^2 + 2\,\vec{k}\vec{q}_i)\,T_{\text{pot}}^{N} \frac{1}{(k+q_i)^2}\right)\,,\notag\\
  ={}& \frac{1}{\vec{q}_i^{\,2}}\left( 1 + (k_0 + q_0)^2  T_{\text{us}}^{M-2} T_{\text{pot}}^{N-2} \frac{1}{(k+q_i)^2}  \right.\notag\\
  &\left. -\vec{k}^2 T_{\text{us}}^{M-2} T_{\text{pot}}^{N} \frac{1}{(k+q_i)^2} - 2\,\vec{k}\vec{q}_i\,T_{\text{us}}^{M-1}T_{\text{pot}}^{N} \frac{1}{(k+q_i)^2}\right)\notag\\
  ={}& T_{\text{pot}}^N T_{\text{us}}^M \frac{1}{(k+q_i)^2}\,,
\end{align}
%----------------------------------------------------------------------------------------------------------------
where we have used the induction hypothesis
%----------------------------------------------------------------------------------------------------------------
\begin{equation}
  \label{eq:induction_hypothesis}
T_{\text{pot}}^n T_{\text{us}}^m \frac{1}{(k+q_i)^2}= T_{\text{us}}^m
T_{\text{pot}}^n\frac{1}{(k+q_i)^2}\quad \text{for }n+m < N+M
\end{equation}
%----------------------------------------------------------------------------------------------------------------
in the last step.

Furthermore, we find
%----------------------------------------------------------------------------------------------------------------
\begin{equation}
  \label{eq:TT_pot_denom}
  T_{\text{pot}}^N T_{\text{us}}^N \frac{1}{(k+q_i)^2} = \sum_{n=0}^N \frac{T_{\text{pot}}^N T_{\text{us}}^N 
\big[(k_0+q_{i0})^2 - 2\vec{q}_i\vec{k} - \vec{k}^2\big]^n}{(q_i^2)^{n+1}}\,,
\end{equation}
%----------------------------------------------------------------------------------------------------------------
where the numerators are simply polynomials in the components of $k$ and $q_i$.

Finally, the exponential (\ref{eq:I}) can be expanded by observing 
%----------------------------------------------------------------------------------------------------------------
\begin{equation}
\vec{x}(t) \sim R,
\end{equation}
%----------------------------------------------------------------------------------------------------------------
i.e. 
%----------------------------------------------------------------------------------------------------------------
\begin{equation}
\vec{k}\vec{x}(t) \sim 1 
\end{equation}
%----------------------------------------------------------------------------------------------------------------
in the potential region and 
%----------------------------------------------------------------------------------------------------------------
\begin{equation}
\vec{k}\vec{x}(t) \sim v 
\end{equation}
%----------------------------------------------------------------------------------------------------------------
in the ultrasoft region. This yields
%----------------------------------------------------------------------------------------------------------------
\begin{align}
  \label{eq:TT_exp}
  T_{\text{pot}}^N T_{\text{us}}^N  \exp\big(i k[x(t_1) - x(t_2)]\big) 
={}& T_{\text{pot}}^N \exp\big(-i k_0 c (t_1-t_2)\big) \sum_{n=0}^N \frac{\big(i \vec{k}[\vec{x}(t_1) 
- \vec{x}(t_2)]\big)^n}{n!}\notag\\
={}& \exp\big(-i k_0 c (t_1-t_2)\big) \sum_{n=0}^N \frac{\big(i \vec{k}[\vec{x}(t_1) 
- \vec{x}(t_2)]\big)^n}{n!}\,,\\
  T_{\text{us}}^N T_{\text{pot}}^N  \exp\big(i k[x(t_1) - x(t_2)]\big) 
={}& T_{\text{us}}^N \exp\big(i k[x(t_1) - x(t_2)]\big) \notag\\
  ={}& \exp\big(-i k_0 c (t_1-t_2)\big) \sum_{n=0}^N 
\frac{\big(i \vec{k}[\vec{x}(t_1) - \vec{x}(t_2)]\big)^n}{n!}\,.
\end{align}
%----------------------------------------------------------------------------------------------------------------
We have now shown that the Taylor expansions commute for each of the
factors in Eq.~\eqref{eq:I} and that the product of all expanded
factors Eqs.~(\ref{eq:TT_J}--\ref{eq:TT_k+p}),
\eqref{eq:TT_pot_denom}, \eqref{eq:TT_exp} has the required form Eq.~\eqref{eq:T_commute}.
%----------------------------------------------------------------------------------------------------------------
%----------------------------------------------------------------------------------------------------------------
\section{The eccentricity expansion of the non--local terms}
\label{sec:B}
%----------------------------------------------------------------------------------------------------------------

\vspace*{1mm}
\noindent
We have recalculated the eccentricity expansion of the non--local terms using standard representations given in 
\cite{Damour:2015isa,Bini:2020wpo,Bini:2020hmy}. To $O(e_t^{20})$ we obtain
%----------------------------------------------------------------------------------------------------------------
\begin{eqnarray}
F_{\rm 4PN}(a_r,e_t) &=&    
\frac{\nu^2}{a_r^5} \Biggl\{
        -\frac{32}{5} (\ln(a_r) - 2\gamma_E)
        +\frac{128}{5} \ln(2)
        + e_t^2 \Biggl[
                -\frac{176}{5}
                -\frac{628}{15} (\ln(a_r) - 2\gamma_E)
\nonumber\\ &&            
    +\frac{296}{15} \ln(2) 
                +\frac{729}{5} \ln(3)
          \Biggr]
        + e_t^4 \Biggl[
                -\frac{2681}{15}
                -121 (\ln(a_r) - 2\gamma_E)
                +\frac{29966}{15} \ln(2)
\nonumber\\ &&            
    -\frac{13851}{20} \ln(3)
          \Biggr]
        + e_t^6 \Biggl[
                -\frac{90017}{180}
                -\frac{763}{3} (\ln(a_r) - 2\gamma_E)
                -\frac{116722}{15} \ln(2)
\nonumber\\ &&             
   +\frac{419661}{320} \ln(3)
                +\frac{1953125}{576} \ln(5)
          \Biggr]
        + e_t^8 \Biggl[
                -\frac{306433}{288}
                -\frac{3605}{8} (\ln(a_r) - 2\gamma_E)
\nonumber\\ &&               
 +\frac{5381201}{180} \ln(2)
                +\frac{26915409}{2560} \ln(3)
                -\frac{83984375}{4608} \ln(5)
          \Biggr]
        + e_t^{10} \Biggl[
                -\frac{18541327}{9600}
\nonumber\\ &&               
 -\frac{114807}{160} (\ln(a_r) - 2\gamma_E)
                -\frac{4697998651}{54000} \ln(2)
                -\frac{138733913079}{2048000} \ln(3)
\nonumber\\ &&        
        +\frac{18736328125}{442368} \ln(5)
                +\frac{678223072849}{18432000} \ln(7)
          \Biggr]
+e_t^{12} \Biggl[
        -\frac{364045577}{115200}-\frac{679679}{640} 
\nonumber\\ &&
\times \big(
                -2 \gamma_E
                +\ln \big(
                        a_r\big)
        \big)
+\frac{110301092701}{216000} \ln(2)
+\frac{1437894581679}{8192000} \ln(3)
\nonumber\\ &&
-\frac{100439453125}{1769472} \ln(5)
-\frac{678223072849}{2949120} \ln(7) \Biggr]
+e_t^{14} \Biggl[
        -\frac{775035553}{161280}-\frac{95381}{64} 
\nonumber\\ && \times
\big(
                -2 \gamma_E
                +\ln \big(
                        a_r\big)
        \big)
-\frac{38217199661503}{15876000} \ln(2) +\frac{996383367472131}{3211264000} \ln(3) 
\nonumber\\ &&
+\frac{155008544921875 
}{3121348608} \ln(5) +\frac{4122918059849071}{6370099200} \ln(7)\Biggr]
+e_t^{16} \Biggl[
        -
        \frac{992166951}{143360}
\nonumber\\ &&
-\frac{1028313}{512} \big(
                -2 \gamma_E
                +\ln \big(
                        a_r\big)
        \big)
+\frac{2885944821108703}{381024000} \ln(2)
\nonumber\\ &&
-\frac{199179784215931689}{51380224000} \ln(3) +\frac{35973968505859375}{49941577728} \ln(5) 
\nonumber\\ &&
-\frac{558878193388980017}{509607936000} \ln(7)\Biggr]
+e_t^{18} \Biggl[
        -\frac{13153280515}{1376256}-\frac{21477885}{8192}
\big(
                -2 \gamma_E
\nonumber\\ &&       
         +\ln \big(
                        a_r\big)
        \big)
-\frac{2994261720973969703}{164602368000} \ln(2)
+\frac{11141707388359168251}{822083584000} \ln(3)
\nonumber\\ &&
-\frac{660797547149658203125}{115065395085312} \ln(5)
+\frac{1113553185418308608323}{880602513408000} \ln(7)
\nonumber\\ &&
+\frac{81402749386839761113321}{43149523156992000} \ln(11) \Biggr]
+e_t^{20} \Biggl[
        -\frac{30115692673}{2359296}-\frac{1639570933}{491520} 
\nonumber\\ && \times
\big(
                -2 \gamma_E
                +\ln \big(
                        a_r\big)\big)
	+\frac{321972869963109745379}{7054387200000} \ln(2) 
\nonumber\\ &&
-\frac{1923804568946611360809}{82208358400000} \ln(3) 
+\frac{85042155284881591796875}{4142354223071232} \ln(5)
\nonumber\\ &&
-\frac{31245295013617800187583}{29353417113600000} \ln(7)
-\frac{13268648150054881061471323}{862990463139840000}\ln(11)\Biggr]
\Biggr\}  
\nonumber\\
&& + O(e_t^{22}), 
\end{eqnarray}
\begin{eqnarray}
\lefteqn{
F_{\rm 5PN}(a_r,e_t) =} \nonumber\\ &&
\frac{\nu^2}{a_r^6} \Biggl\{
         \frac{2}{105} (2927+588 \nu ) (\ln(a_r) - 2 \gamma_E)
        +\frac{4}{105} (-6319+684 \nu ) \ln(2)
\nonumber\\ &&       
 -\frac{243}{14} (-1+4 \nu ) \ln(3)
        -\frac{96}{5}
        +\frac{32 \nu }{5}
%---        
        + e_t^2 \Biggl[
                 \frac{2}{105} (8004+11935 \nu ) (\ln(a_r) - 2 \gamma_E)
\nonumber\\ &&             
   -\frac{4}{105} (-14268+100247 \nu ) \ln(2)
                +\frac{729}{70} (-76+129 \nu ) \ln(3)
                -\frac{5441}{35}
                +\frac{4672 \nu }{21}
        \Biggr]
%---        
\nonumber\\ &&      
  + e_t^4 \Biggl[
         	 \frac{1}{70} (-14003+78435 \nu ) (\ln(a_r) - 2\gamma_E)
                +\frac{1}{105} (-599911+3476231 \nu ) \ln(2)
\nonumber\\ &&
                -\frac{729 (2085+34396 \nu )}{4480} \ln(3)
                -\frac{9765625 (-1+4 \nu )}{2688} \ln(5)
                -\frac{1160639}{840}
                +\frac{59756 \nu }{35}
        \Biggr]
%---  
\nonumber\\ &&      
        + e_t^6 \Biggl[
                 \frac{1}{30} (-53391+100135 \nu ) (\ln(a_r) -  2 \gamma_E)
                -\frac{73}{945} (-226057+3011889 \nu ) \ln(2)
\nonumber\\ &&             
   -\frac{243}{896} (-95283+292001 \nu ) \ln(3)
                +\frac{78125 (-5557+43575 \nu )}{24192} \ln(5)
                -\frac{15761437}{2520}
\nonumber\\ &&              
  +\frac{474653 \nu }{72}
        \Biggr]
%---        
        + e_t^8 \Biggl[
                 \frac{1}{64} (-356481+490280 \nu ) (\ln(a_r) - 2 \gamma_E)
\nonumber\\ &&       
         +\frac{(-1814239887+33331273432 \nu )}{30240} \ln(2)
                +\frac{729 (-181288681+1304180292 \nu )}{1146880} 
\nonumber\\ && \times
\ln(3)
                -\frac{78125 (325441+43174300 \nu )}{6193152} \ln(5)
                -\frac{96889010407 (-1+4 \nu )}{884736} \ln(7)
\nonumber\\ &&              
  -\frac{168508293}{8960}
                +\frac{2591779 \nu }{144}
        \Biggr]
%---        
        + e_t^{10} \Biggl[
                +\frac{561}{320} (-7258+8547 \nu )(\ln(a_r) - 2\gamma_E)  
\nonumber\\ &&           
     +\frac{(1135478771202-6934343243023 \nu )}{756000} \ln(2)
\nonumber\\ &&
                -\frac{2187 (-1924018874+43692700941 \nu )}{28672000} \ln(3)
                +\frac{1953125 (69962+1244683 \nu )}{2064384} 
\nonumber\\ && \times
\ln(5)
                +\frac{282475249 (-244698+1611757 \nu )}{110592000} \ln(7)
                -\frac{1492974817}{33600}
                +\frac{255777929 \nu }{6400}
        \Biggr]
\nonumber\\ &&
+ e_t^{12} \Biggl[
+ \Biggl(-\frac{3926923}{160} + \frac{6733727 \nu}{256} \Biggr)( \ln(a_r) - 2 \gamma_E) 
+ \Biggl(-\frac{30140932254133}{3402000} 
\nonumber\\ &&
+ \frac{109595282746879 \nu}{1814400} \Biggr) \ln(2) 
+ \Biggl(\frac{431854060307859}{131072000} 
- \frac{37854312670341 \nu}{6553600} \Biggr)  \ln(3) 
\nonumber\\ &&
- \Biggl(\frac{47376871796875}{891813888} 
+ \frac{60736572265625 \nu}{37158912} \Biggr)  \ln(5) 
+ \Biggl(\frac{30496744521746713}{21233664000} 
\nonumber\\ &&
- \frac{90299298142188709 \nu}{5308416000}\Biggr) \ln(7) 
-\frac{20691354791}{230400} 
+ \frac{3566474429 \nu}{46080} \Biggr] 
\nonumber\\ &&
%--
+ e_t^{14} \Biggl[
\Biggl(-\frac{190225893}{4480} 
+ \frac{5466461 \nu}{128} \Biggr)(\ln(a_r) - 2 \gamma_E)
+ \Biggl(\frac{4290237684292667}{133358400} 
\nonumber\\ &&
- \frac{19381110496948853 \nu}{74088000}\Biggr) \ln(2) 
+ \Biggl(-\frac{217154530377393003}{8991539200} 
\nonumber\\ &&
+ \frac{5487909996792714903 \nu}{44957696000} \Biggr) \ln(3)
+ \Biggl(\frac{248411703945390625}{43698880512} 
\nonumber\\ &&
- \frac{34494852294921875 \nu}{1618477056} \Biggr) \ln(5) 
+ \Biggl(-\frac{5548114000135259}{2359296000} 
\nonumber\\ &&
+ \frac{295225368493618129 \nu}{7077888000}\Biggr) \ln(7)
-\frac{122529921403}{752640} + \frac{6267828367 \nu}{46080} \Biggr]
\nonumber\\ &&
+e_t^{16} \Biggl[
 \Biggl(-\frac{39012948117}{573440} + \frac{267733323 \nu}{4096}\Biggr)(\ln(a_r) - 2\gamma_E)
\nonumber\\ &&
-\frac{8748776901657}{32112640} + \frac{51000706143}{229376} \nu
+\Biggl(-\frac{569049681664742359}{5689958400} 
\nonumber\\ &&
+ \frac{17916377627561875829}{21337344000} \nu \Biggr) \ln(2)
+ \Biggl(\frac{192054917627699907573}{2301834035200} 
\nonumber\\ &&
- \frac{1707020784493974596637}{2877292544000} \nu
\Biggr) \ln(3)
+ \Biggl(-\frac{8138674822699358046875}{178990614577152} 
\nonumber\\ &&
+ \frac{214230939178466796875}{913217421312} \nu
\Biggr) \ln(5)
+ \Biggl(\frac{28987110744247081153}{7247757312000} 
\nonumber\\ &&
- \frac{224384811477454209253}{3261490790400} \nu
\Biggr) \ln(7)
+ \Biggl(\frac{81402749386839761113321}{4474765364428800} -
\nonumber\\ &&
\frac{81402749386839761113321}{1118691341107200} \nu \Biggr) \ln(11)
\Biggr]
%---
+ e_t^{18} \Biggl[
-\frac{30986035007243}{72253440}
\nonumber\\ &&
+ \frac{945865602119}{2752512} \nu
     +\Biggl(-\frac{4426707571}{43008} 
+ \frac{1565114265}{16384} \nu \Biggr)(\ln(a_r) - 2 \gamma_E)
\nonumber\\ &&
+ \Biggl(\frac{328393145609363098807}{864162432000}
  - \frac{862890935509435128773}{329204736000} \nu \Biggr) \ln(2)
\nonumber\\ &&
+ \Biggl(-\frac{203719103345458519569}{1438646272000}
  + \frac{16662707951693189189427}{11509170176000} \nu \Biggr) \ln(3)
\nonumber\\ &&
+ \Biggl(\frac{106107911818692516640625}{604093324197888}
  - \frac{772369312821197509765625}{690392370511872} \nu \Biggr) \ln(5)
\nonumber\\ &&
+ \Biggl(-\frac{4145697650045386774577}{660451885056000}
   + \frac{144186209646215652673877}{1761205026816000} \nu \Biggr) \ln(7)
\nonumber\\ &&
- \Biggl(\frac{35656966913567027996954261}{226534996574208000}
  - \frac{470263683207773299951655417}{604093324197888000} 
\nonumber\\ &&
\nu \Biggr) \ln(11)
\Biggr]
%---
  + e_t^{20} \Biggl[
  -\frac{177047266722689}{275251200}
+ \frac{11982611767181}{23592960} \nu
\nonumber\\ &&
+  \Biggl(-\frac{48797431397}{327680} + \frac{132236197093}{983040}\nu \Biggr) (\ln(a_r) - 2 \gamma_E)
\nonumber\\ &&
+ \Biggl(-\frac{1113890589789636783281623}{691329945600000}
  + \frac{6591498923973012595615229}{691329945600000} \nu \Biggr) \ln(2)
\nonumber\\ &&
+ \Biggl(-\frac{21964236463404827739962163}{147317378252800000}
  - \frac{27978900275432534562921069}{36829344563200000} \nu \Biggr) \ln(3)
\nonumber\\ &&
- \Biggl(\frac{420710016768367311936953125}{927887345967955968}
  - \frac{760960337767734527587890625}{231971836491988992} \nu \Biggr) \ln(5)
\nonumber\\ &&
+ \Biggl(\frac{31701323997378511368405793}{4226892064358400000}
  - \frac{77737887155913025713859691}{1056723016089600000} \nu \Biggr) \ln(7)
\nonumber\\ &&
+ \Biggl(\frac{7854405708884154695671648813}{11835297780203520000}
  - \frac{112327084059654478481943854653}{28996479561498624000} \nu \Biggr) \ln(11)
\nonumber\\ &&
+ \Biggl(\frac{91733330193268616658399616009}{579929591229972480000}
- \frac{91733330193268616658399616009}{144982397807493120000} \nu\Biggr) \ln(13) \Biggr]
\Biggr\} 
\nonumber\\ &&
+ O(e_t^{22}).
\end{eqnarray}
%----------------------------------------------------------------------------------------------------------------
%----------------------------------------------------------------------------------------------------------------
\section{The contour integral for the Delaunay variable \boldmath $i_r$}
\label{sec:C}
%----------------------------------------------------------------------------------------------------------------

\vspace*{1mm}
\noindent
The integral $J_1$, describing effect of the Newton dynamics, is given by
%----------------------------------------------------------------------------------------------------------------
\begin{eqnarray}
J_1 = \frac{1}{2 \pi i} \oint dx \sqrt{A + \frac{2B}{x} + \frac{C}{x^2}} &=& \frac{B}{\sqrt{-A}} - \sqrt{-C}.
\end{eqnarray}
%----------------------------------------------------------------------------------------------------------------
All the remaining integrals are directly obtained from the residue at $x=0$.

The integral in (\ref{eq:INTperi}) reads to 5PN
%----------------------------------------------------------------------------------------------------------------
\begin{eqnarray}
i_r \hspace*{-2mm} &=&  \hspace*{-2mm} \frac{B}{\sqrt{-A}} - \sqrt{-C} \Biggl\{
1
      - \eta^2 \frac{B D_1}{2C^2} 
      + \eta^4 \Biggl[
        \frac{3 D_1^2 \big(
                -5 B^2
                +A C
        \big)}{16 C^4}
        -\frac{D_2 \big(
                -3 B^2
                +A C
        \big)}{4 C^3}
 +\frac{B D_3}{4 C^4} 
\nonumber\\ &&
\times \big(-5 B^2
 	               +3 A C
        \big)
\Biggr]
+ \eta^6 \Biggl[
        -\frac{B D_5 \big(
                63 B^4
                -70 A B^2 C
                +15 A^2 C^2
        \big)}{16 C^6}
                +        D_1 \Biggl(
\frac{5 B D_2 \big(
                        7 B^2
                        -3 A C
                \big)}{8 C^5} 
\nonumber\\ && 
                -\frac{15 D_3 \big(
                        21 B^4
                        -14 A B^2 C
                        +A^2 C^2
                \big)}{32 C^6} \Biggr)
        +\frac{35 B D_1^3 \big(
                -3 B^2
                +A C
        \big)}{32 C^6}
        +\frac{D_4}{16 C^5} \big(
                35 B^4
                -30 A B^2 C
\nonumber\\ &&
                +3 A^2 C^2
        \big)
\Biggr]
+ \eta^8 \Biggl[
        \frac{7 D_3^2}{128 C^8} \big(
                -429 B^6
 +495 A B^4 C
                -135 A^2 B^2 C^2
\nonumber\\ &&              
  +5 A^3 C^3
        \big)
        -\frac{D_6}{32 C^7} \big(
                -231 B^6
                +315 A B^4 C
                -105 A^2 B^2 C^2
                +5 A^3 C^3
        \big)
        +\frac{B D_7}{32 C^8} \big(
                -429 B^6
\nonumber\\ &&              
  +693 A B^4 C
                -315 A^2 B^2 C^2
                +35 A^3 C^3
        \big)
        + D_1
         \Biggl(
                \frac{7 D_5}{64 C^8} \big(
                        -429 B^6
 +495 A B^4 C 
\nonumber\\ &&
                        -135 A^2 B^2 C^2
                        +5 A^3 C^3
                \big)
                +\frac{21 B D_4 \big(
                        33 B^4
                        -30 A B^2 C
                        +5 A^2 C^2
                \big)}{32 C^7}
        \Biggr)
\nonumber\\ &&
        + D_1^2 \Biggl(
                \frac{105 D_2 \big(
                        33 B^4
                        -18 A B^2 C
                        +A^2 C^2
                \big)}{128 C^7}
 -\frac{63 B D_3 \big(
                        143 B^4
                        -110 A B^2 C
                        +15 A^2 C^2
                \big)}{128 C^8}
        \Biggr)
\nonumber\\ &&               
        -\frac{105 D_1^4 \big(
                143 B^4
                -66 A B^2 C
                +3 A^2 C^2
        \big)}{1024 C^8}
        -\frac{15 D_2^2 \big(
                21 B^4
                -14 A B^2 C
                +A^2 C^2
        \big)}{64 C^6}
\nonumber\\ &&      
  +\frac{21 B D_2 D_3 \big(
                33 B^4
                -30 A B^2 C
                +5 A^2 C^2
        \big)}{32 C^7}
\Biggr]
+\eta ^{10} \Biggl[
        -\frac{D_8}{256 C^9} \big(
                6435 B^8
                -12012 A B^6 C
\nonumber\\ &&               
                +6930 A^2 B^4 C^2
 -1260 A^3 B^2 C^3
                +35 A^4 C^4
        \big)
        +\frac{B D_9}{256 C^{10}} \big(
                12155 B^8
                -25740 A B^6 C
\nonumber\\ &&              
  +18018 A^2 B^4 C^2
                -4620 A^3 B^2 C^3
                +315 A^4 C^4
        \big)
        +D_2 \Biggl(
                -\frac{9 B D_5}{64 C^9} \big(
                        715 B^6
          -1001 A B^4 C
\nonumber\\ &&              
                        +385 A^2 B^2 C^2
                        -35 A^3 C^3
                \big)
                +\frac{7 D_4}{64 C^8} \big(
                        429 B^6
                        -495 A B^4 C
                        +135 A^2 B^2 C^2
                        -5 A^3 C^3 \big)
                \Biggr)
        +D_1^2 \Biggl(
\nonumber\\ &&
                \frac{495 B D_5}{256 C^{10}} \big(
                        221 B^6
                        -273 A B^4 C
                        +91 A^2 B^2 C^2
                        -7 A^3 C^3
                \big)
                -\frac{315 D_4}{256 C^9} \big(
                        143 B^6
                        -143 A B^4 C
\nonumber\\ &&                     
   +33 A^2 B^2 C^2
                        -A^3 C^3
                \big)
        \Biggr)
        +D_3 \Biggl(
                -\frac{9 B D_4}{64 C^9} \big(
                        715 B^6
                        -1001 A B^4 C
 +385 A^2 B^2 C^2
                        -35 A^3 C^3
                \big)
\nonumber\\ &&                       
 +\frac{45 D_5}{512 C^{10}} \big(
                        2431 B^8
                        -4004 A B^6 C
                        +2002 A^2 B^4 C^2
                        -308 A^3 B^2 C^3
+7 A^4 C^4
                \big)
        \Biggr)
        +D_1^3 \Biggl(
                \frac{1155 D_3}{512 C^{10}} 
\nonumber\\ &&                        
\big(
                        221 B^6
                        -195 A B^4 C
                        +39 A^2 B^2 C^2
                        -A^3 C^3
                \big)
                -\frac{1155 B \big(
                        39 B^4
                        -26 A B^2 C
                        +3 A^2 C^2
                \big) D_2}{256 C^9}
        \Biggr)
\nonumber\\ &&
        +D_1 \Biggl(
                -
                \frac{9 B D_6}{64 C^9} \big(
                        715 B^6
                        -1001 A B^4 C
                        +385 A^2 B^2 C^2
                        -35 A^3 C^3
                \big)
                +\frac{495 B D_3^2}{256 C^{10}} \big(
                        221 B^6
\nonumber\\ &&                       
 -273 A B^4 C
                        +91 A^2 B^2 C^2
                        -7 A^3 C^3
                \big)
                -\frac{315 D_2 D_3}{128 C^9} \big(
                        143 B^6
                        -143 A B^4 C
                        +33 A^2 B^2 C^2
                        -A^3 C^3
                \big)
\nonumber\\ &&               
 +\frac{45 D_7}{512 C^{10}} \big(
                        2431 B^8
                        -4004 A B^6 C
                        +2002 A^2 B^4 C^2
                        -308 A^3 B^2 C^3
                        +7 A^4 C^4
                \big)
\nonumber\\ &&
                +\frac{63 B \big(
                        143 B^4
                        -110 A B^2 C
                        +15 A^2 C^2
                \big) D_2^2}{128 C^8}
        \Biggr)
        +\frac{9009 B \big(
                17 B^4
                -10 A B^2 C
                +A^2 C^2
        \big) D_1^5}{2048 C^{10}}
\Biggr]
\Biggr\}.
\end{eqnarray}
%----------------------------------------------------------------------------------------------------------------
The coefficients $A$ to $D_9$ are determined iteratively expanding (\ref{eq:RES}) in powers of $\eta^2$. They depend on 
the respective Hamiltonian for which one may choose a pole- and log--free form.

\vspace*{5mm}
\noindent
{\bf Acknowledgment.} 
We thank D. Bini, Th.~Damour, S.~Foffa, C.~Kavanagh, K.~Sch\"onwald, V.~Smirnov, and B.~Wardell for discussions 
and L.~Blanchet for a remark. This work has been funded in part by EU TMR network SAGEX agreement No. 764850 
(Marie Sk\l{}odowska--Curie) and COST action CA16201: Unraveling new physics at the LHC through the precision 
frontier. G.~Sch\"afer has been supported in part by Kolleg Mathematik Physik Berlin (KMPB) and DESY. Part of 
the text has been typesetted using {\tt SigmaToTeX} of the package {\tt Sigma} \cite{SIG1,SIG2}.
%-------------------------------------------------------------------------------------

%-------------------------------------------------------------------------------------
\end{document}